\documentclass[useAMS,usenatbib,twocolumn]{mn2e}
\usepackage{amssymb}
\usepackage{amsmath}
\usepackage{epsfig}
\usepackage{subfigure}

\begin{document}

\title[Inhomogeneous Gravity]{ Inhomogeneous Gravity}
\author[T. Clifton, D. F. Mota and J. D. Barrow]{Timothy Clifton$^1$\thanks{
E-mail: T.Clifton@damtp.cam.ac.uk}, David F. Mota$^{2}$\thanks{
E-mail: D.F.Mota@damtp.cam.ac.uk} and John D. Barrow$^1$\thanks{
E-mail: J.D.Barrow@damtp.cam.ac.uk} \\
%EndAName
$^1$ Department of Applied Mathematics and Theoretical Physics, Centre for
Mathematical Sciences, \\
University of Cambridge, Wilberforce Road, Cambridge CB3 0WA, UK\\
$^2$Astrophysics, Department of Physics, University of Oxford, Keble Road,
Oxford, OX1 3RH, UK}

\date{\today }
%\pacs{xxx.xx}
\maketitle
\begin{abstract}
We study the inhomogeneous cosmological evolution of the Newtonian
gravitational 'constant' $G$ in the framework of scalar-tensor theories. We
investigate the differences that arise between the evolution of $G$ in the
background universes and in local inhomogeneities that have separated out
from the global expansion. Exact inhomogeneous solutions are found which
describe the effects of masses embedded in an expanding FRW Brans-Dicke
universe. These are used to discuss possible spatial variations of $G$ in
different regions. We develop the technique of matching different
scalar-tensor cosmologies of different spatial curvature at a boundary. This
provides a model for the linear and non-linear evolution of spherical
overdensities and inhomogeneities in $G.$ This allows us to compare the
evolution of $G$ and $\dot{G}$ that occurs inside a collapsing overdense
cluster with that in the background universe. We develop a simple
virialisation criterion and apply the method to a realistic lambda-CDM
cosmology containing spherical overdensities. Typically, far slower
evolution of $\dot{G}$ will be found in the bound virialised cluster than in
the cosmological background. We consider the behaviour that occurs in
Brans-Dicke theory and in some other representative scalar-tensor theories.
\end{abstract}
\pagerange{\pageref{firstpage}--\pageref{lastpage}} \pubyear{2004}
\label{firstpage}
%
%\begin{pacs}
%98.80.-k  06.20.Jr
%\end{pacs}

%\maketitle

\section{Introduction}

Many of the most interesting extensions of the general theory of relativity
are scalar-tensor theories of gravity. They incorporate scalar fields which
can carry space-time variations in scalar quantities that are traditionally
assumed to be constant in general relativity. In this respect they provide
natural arenas in which to explore the consequences of varying 'constants'
of Nature, like Newton's 'constant' $G$, or Sommerfeld's fine structure
'constant', $\alpha .$ In this paper we shall focus upon the first of these
applications, pioneered by Jordan \citep{Jordan49,Jordan491,Jordan492} and then refined into the
most familiar generalisation of Einstein's theory of gravitation by Brans
and Dicke in 1961, \citep{Brans61}. This theory features both in direct
explorations of the possible variation of $G$, in dimensional reduction of
higher-dimensional theories, and in string theories \citep{Green}, where is
appears containing a dilaton with non-minimal coupling. The original theory
of Brans-Dicke is recognised as the simplest case of a family of
scalar-tensor gravity theories in which the original Brans-Dicke coupling
parameter, $\omega $, becomes a function of the scalar field that carries
variations in $G$, \citep{Bergmann68},\citep{Wagoner70, nor}.

Extensive studies have been made of cosmological solutions of scalar-tensor
gravity theories \citep{Fuji}, although they are limited in two respects.
First, they focus on the simplest case of isotropic expansion with zero
spatial curvature, where simple exact solutions exist. Second, they are
exclusively concerned with spatially homogeneous cosmologies. The latter
restriction means that the value of $G$ and its rate of change in time, $%
\dot{G},$ are required to be the same everywhere in the universe. This
assumption has run through the entire literature on varying $G$ and so it is
generally assumed, for example, that local observational bounds on varying $%
G $ derived from geophysics, solar system dynamics, stellar evolution, or
white dwarf cooling can be applied directly to constrain cosmological
variations in $G$ on extragalactic scales or in the very early universe.
There is no justification for this simplifying assumption, as pointed
out by Barrow and O'Toole \citep{Barrow01}. The local bounds
on varying $G$ are all derived from observations of gravitationally bound
'lumps' which are in gravitational equilibrium and \textit{do not take part
in the expansion of the universe. }Before they can be extrapolated to
constrain possible variations of $G$ in a background Friedmann universe (as
is habitually done in the literature, without justification) we need to
understand how $G$ and $\dot{G}$ are expected to vary in space in a
realistic inhomogeneous universe. Since, even on the scale of a typical
galaxy, the amplitude of visible density inhomogeneities are of order $%
10^{6} $, we need to go beyond linear perturbation theory for such an
analysis.

In this paper we begin to confront this deficiency by tracing the evolution
of $G$, first in exact inhomogeneous solutions and then a simple, but not
unrealistic, inhomogeneous universe in which a zero-curvature
Brans-Dicke-Friedmann background universe is populated by spherical
overdensities which are modelled by positive curvature Brans-Dicke-Friedmann
universes in the dust-dominated era of the universe's history. This will
enable us to track the different evolution followed by $G(t)$ in the
background universe and in the overdense regions, which eventually separate
off from the background universe and start to contract to high-density like
separate closed universes. This process can produce significant differences
between $G$ and $\dot{G}$ in the background and in the overdensities.
Eventually, the collapse of the spherical overdensities will be stopped by
pressure and a complicated sequence of dissipative and relaxation processes
will lead to virialisation and some final state of gravitational
equilibrium. This state will provide the gravitational environment out of
which which stars and planetary systems like our own will form, directly
reflecting the local value of $G(t)$ inherited from their virialised
protogalaxy or its parent protocluster. The simple model we use for
inhomogeneities in density and in $G$ has many obvious limitations, notably
in its neglect of pressure, deviations from spherical symmetry, accretion,
and interactions between inhomogeneities. Nonetheless, we expect that it
will be indicative of the importance of taking spatial inhomogeneity into
account in any attempts to use observational data to constrain cosmological
models which permit varying $G$. It provides the first step in a clear path
towards improved realism in the modelling of inhomogeneities that mirrors
the route followed in standard cosmological studies of galaxy formation with
constant $G$.

In section \ref{Scalar-Tensor Cosmology} we give the field equations and the
field equations for scalar-tensor gravity theories. To fix ideas, we use
some exact Brans--Dicke--Friedmann cosmological solutions in section \ref%
{Analytic} to model the time variation of $G$ in some new exact
inhomogeneous solutions of the field equations which describe a spherical
inhomogeneity in an expanding universe. In section \ref{vaccollapse} we use
exact solutions of flat and closed vacuum Brans--Dicke--Friedmann
cosmologies to illustrate the use of the spherical collapse model for
non-linear overdensities. We then apply the same techniques to dust--filled
Brans--Dicke--Friedmann universes in Section \ref{Results}. The numerical
solution of the equations are presented and discussed in this section for
Brans-Dicke and some other scalar--tensor theories. A summary of our
principal results is given in \ref{Conclusions}.

\section{Scalar-Tensor Cosmologies}

\label{Scalar-Tensor Cosmology}

The action for a scalar-tensor theory of gravity is given by

\begin{equation}
S=\frac{1}{16\pi }\int \,d^{4}x\sqrt{-g}(\phi R+\frac{\omega (\phi )}{\phi }%
g^{\mu \nu }\partial _{\mu }\phi \partial _{\nu }\phi +16\pi \mathcal{L}%
_{m}).  \label{JBDaction}
\end{equation}%
Here $\phi $ is a scalar field, $R$ is the Ricci scalar, and $L_{m}$ is the
Lagrangian for matter fields in the space-time and the free function $\omega
(\phi )$ specifies the scalar-tensor theory. In Brans-Dicke theory, $\omega $
is constant. The action above is varied with respect to $g_{\mu \nu }$ to
give the field equations

\begin{multline}
R^{\mu \nu }-\frac{1}{2}g^{\mu \nu }R+\frac{1}{\phi }(g^{\mu \rho }g^{\nu
\sigma }-g^{\mu \nu }g^{\rho \sigma })\phi _{;\rho \sigma }
\\
+\frac{\omega(\phi )}{\phi ^{2}}(g^{\mu \rho }g^{\nu \sigma }-\frac{1}{2}g^{\mu \nu
}g^{\rho \sigma })\phi _{,\rho }\phi _{,\sigma }=-\frac{8\pi }{\phi }T^{\mu
\nu },
\label{field equations}
\end{multline}%
and with respect to $\phi $ to obtain the propagation equation

\begin{equation}
\square \phi =\frac{1}{2\omega (\phi )+3}(8\pi T-\omega ^{\prime }(\phi
)g^{ab}\phi _{;a}\phi _{;b}),  \label{box phi}
\end{equation}%
where prime denotes differentiation with respect to $\phi $. In this paper
we will often consider Friedmann-Robertson-Walker (FRW) universes with the
metric

\begin{equation*}
ds^{2}=dt^{2}-a^{2}(t)\left( \frac{dr^{2}}{1-kr^{2}}+r^{2}(d\theta ^{2}+\sin
^{2}\theta d\phi ^{2})\right) ,
\end{equation*}%
where $a(t)$ is the scale factor and $k$ is the curvature parameter. Using
the FRW metric in eq. (\ref{field equations}) gives

\begin{equation}
2\dot{H}+3H^{2}+\frac{\omega }{2}\frac{\dot{\phi}^{2}}{\phi ^{2}}+2H\frac{%
\dot{\phi}}{\phi }+\frac{\ddot{\phi}}{\phi }=-\frac{8\pi }{\phi }p-\frac{k}{%
a^{2}},  \label{Nariai1}
\end{equation}

\begin{equation}
\frac{\ddot{\phi}}{\phi }=\frac{8\pi }{\phi }\frac{(\rho -3p)}{(2\omega +3)}%
-3H\frac{\dot{\phi}}{\phi }-\frac{\dot{\omega}}{(2\omega +3)}\frac{\dot{\phi}%
}{\phi },  \label{Nariai2}
\end{equation}%
and 
\begin{equation}
\frac{8\pi }{3\phi }\rho =H^{2}+H\frac{\dot{\phi}}{\phi }-\frac{\omega }{6}%
\frac{\dot{\phi}^{2}}{\phi ^{2}}+\frac{k}{a^{2}},  \label{Friedmann}
\end{equation}%
where $H\equiv \dot{a}/a$ is the Hubble rate, over-dot denotes
differentiation with respect to comoving proper time, $t$, $\rho $ is the
matter density, and $p$ is the pressure. Each non-interacting fluid source $%
p(\rho )$ separately satisfies a conservation equation:

\begin{equation}
\dot{\rho}+3H(\rho +p)=0.  \label{fluid}
\end{equation}

Substituting eqs. (\ref{Nariai2}) and (\ref{Friedmann}) into eq. (\ref%
{Nariai1}) gives

\begin{multline}
\dot{H}+H^{2}-H\frac{\dot{\phi}}{\phi }+\frac{\omega }{3}\frac{\dot{\phi}^{2}%
}{\phi ^{2}}
\\=-\frac{8\pi }{3\phi }\frac{(3p\omega +3\rho +\rho \omega )}{%
(2\omega +3)}+\frac{1}{2}\frac{\dot{\omega}}{(2\omega +3)}\frac{\dot{\phi}}{%
\phi }.  \label{acceleration}
\end{multline}

%\section{Non-Linear Large Scale Structure Formation}
%\label{Structure Formation}
A general feature of the scalar-tensor field equations is that any solution
of general relativity (hence $\omega $ and $\phi $ both constant) for which
the energy-momentum tensor of matter has vanishing trace (eg vacuum,
black-body radiation, Yang-Mills field, or magnetic field) is a particular ($%
\phi =$ constant) exact solution of the scalar-tensor gravity theory. A
specification of $\omega (\phi )$ is required to determine the theory and
close the system of equations. In general, we do not know the form of $%
\omega (\phi )$ but if the theory is to approach general relativity in an
appropriate limit then we require both $\omega \rightarrow \infty $ and $%
\omega ^{\prime }(\phi )\omega ^{-3}\rightarrow 0$ to hold simultaneously in
the weak--field limit.

\section{Brans-Dicke Cosmologies}

\label{Analytic}

\bigskip \emph{\ }Consider the simplest case of Brans-Dicke (BD) theory \citep%
{Brans61} to fix ideas about varying $G$. In these theories $\omega (\phi
)\equiv \omega $ is a constant. The three essential field equations
for the evolution of the BD scalar field $\phi (t)$ and the expansion scale
factor $a(t)$ in a BD universe are (\ref{Nariai2}), (\ref{Friedmann}), and (%
\ref{fluid}).  Now, $\omega $ is a constant parameter and the theory reduces to
general relativity in the limit $\omega \rightarrow \infty $ where $\phi
=G^{-1}\rightarrow $ constant. The form of the general solutions to the
Friedmann metric in BD theories are fully understood \citep{bar}, \citep{fink,
wands}. The vacuum solution is the $t\rightarrow 0$ attractor for the
perfect-fluid solutions. The general perfect-fluid solutions with equation
of state

\begin{equation}
p=(\gamma -1)\rho  \label{pee}
\end{equation}%
and $k=0$ can all be found. At early times they approach the vacuum
solutions but at late time they approach particular power-law exact
solutions \citep{nar}:

\begin{equation}
a(t)=t^{[2+2\omega (2-\gamma )]/[4+3\omega \gamma (2-\gamma )]}  \label{bds1}
\end{equation}

\begin{equation}
\phi (t)=\phi _{0}t^{[2(4-3\gamma )/[4+3\omega \gamma (2-\gamma )]}
\label{bds2}
\end{equation}

These particular exact power-law solutions for $a(t)$ and $\phi (t)$ are
'Machian' in the sense that the cosmological evolution is driven by the
matter content rather than by the kinetic energy of the free $\phi $ field.
The sign of $\dot{\phi}$ is determined by the sign of $4-3\gamma $.

These solutions are spatially homogeneous and so cannot tell us about the
effects of any spatial inhomogeneity in $\phi $ and $\rho $ on observational
tests of time-varying $G=\phi ^{-1}$. Next, we consider some simple exact
inhomogeneous solutions of BD theory in order to gain some intuition about
the likely effects of spatial inhomogeneity in $G$ . We will find that these
exact simple homogeneous solutions play an important role in determining the
time dependence of $G$ in inhomogeneous solutions.

\subsection{An Inhomogeneous vacuum Brans-Dicke Solution}

\label{vacBD}

It is well known that BD theory is related to general relativity through a
conformal transformation of the form \citep{Fuji}

\begin{equation}
g_{\mu \nu }=\frac{1}{\phi }\bar{g}_{\mu \nu }  \label{conformal}
\end{equation}%
where $\phi $ is the BD scalar field. Symbols with bars refer to quantities
in the Einstein (general relativistic) conformal frame and symbols without
bars refer to quantities in the Jordan (BD) conformal frame. This conformal
equivalence allows us to exploit known solutions of Einstein's field
equations with a scalar field to find solutions to the BD field equations in
a vacuum.

We proceed by first showing explicitly the conformal equivalence of general
relativity with a scalar field and BD in a vacuum under the transformation (%
\ref{conformal}). From (\ref{conformal}) we immediately obtain \citep{Fuji}

\begin{equation}
g^{\mu \nu }=\phi \bar{g}^{\mu \nu },\qquad \sqrt{-g}=\phi
^{-2}\sqrt{-\bar{g}}  \nonumber
\end{equation}
and
\begin{equation}
R=\phi (\bar{R}+6\square \Gamma +6\bar{g}^{\mu \nu
}\Gamma,_{\mu }\Gamma,_{\nu })  \label{R}
\end{equation}
where here $R$ is the Ricci scalar, $\Gamma =\ln \phi ^{-1/2}$ and $\square
\Gamma =\frac{1}{\sqrt{-\bar{g}}}\partial _{\mu }(\sqrt{-\bar{g}}\bar{g}%
^{\mu \nu }\partial _{\nu }\Gamma)$. To derive the Einstein field equations with
a minimally coupled scalar field, we can extremise the Lagrangian density

\begin{equation}
\mathcal{L} = \sqrt{-\bar{g}} \left(\frac{1}{16\pi} \bar{R} + \frac{1}{2} 
\bar{g}^{\mu\nu} \psi,_\mu \psi,_\nu\right)  \label{GR}
\end{equation}
to get Einstein's field equations, $G_{\mu \nu }=-8\pi T_{\mu \nu }$, where $%
T_{\mu \nu }=\psi ,_{\mu }\psi ,_{\nu }-\frac{1}{2}\bar{g}_{\mu \nu }\bar{g}%
^{\alpha \beta }\psi ,_{\alpha }\psi ,_{\beta }$. We now set $\phi =\exp
[\psi \sqrt{\frac{8\pi }{\omega +\frac{3}{2}}}]$ so that under the conformal
transformation prescribed by (\ref{conformal}) we find that (\ref{GR})
becomes

\begin{multline}
\mathcal{L}= \phi^2 \sqrt{-g} (\frac{1}{16\pi} (\phi^{-1} R - 6 \square
\Gamma - 6 \phi^{-1} g^{\mu\nu} \Gamma,_\mu \Gamma,_\nu) \\+ \frac{1}{16\pi}%
(\omega + \frac{3}{2}) \phi^{-1} g^{\mu\nu} \frac{\phi,_\mu \phi,_\nu}{\phi^2%
}).  \label{GR2}
\end{multline}

We see that $\square \Gamma $ disappears on integration by parts and so we
can discard it. We also note that $g^{\mu \nu }\Gamma ,_{\mu }\Gamma ,_{\nu
}=\frac{1}{4}g^{\mu \nu }\phi ^{-2}\phi ,_{\mu }\phi ,_{\nu }$, so that (\ref%
{GR2}) simplifies to

\begin{equation}
\mathcal{L}=\sqrt{-g}\frac{1}{16\pi }(\phi R+\frac{\omega }{\phi }g^{\mu \nu
}\phi ,_{\mu }\phi ,_{\nu }).  \label{JBD}
\end{equation}%
This is the Lagrangian density that gives the BD action (\ref{JBDaction}),
with $\mathcal{L}_{matter}=0$.

Starting with a solution of Einstein's field equations with a scalar field
we can apply this conformal transformation to arrive at a solution of the BD
field equations in a vacuum. A spherically symmetric exact solution for the
collapse of a minimally-coupled scalar field, $\psi $, in general relativity
is known and is given by \citep{Hussain}

\begin{multline}
ds^{2}=(qt+b)(f^{2}(r)dt^{2}-f^{-2}(r)dr^{2})\\-R^{2}(r,t)(d\theta ^{2}+\sin
^{2}\theta d\phi ^{2}),  \label{GRsolution}
\end{multline}%
where $f^{2}(r)=(1-\frac{2c}{r})^{\alpha }$, $R^{2}(r,t)=(qt+b)r^{2}(1-\frac{%
2c}{r})^{1-\alpha }$ and $\alpha =\pm \frac{\sqrt{3}}{2}$. The evolution of
the minimally coupled scalar field, $\psi $, in the Einstein frame, is given
by

\begin{equation}
\psi (r,t)=\pm \frac{1}{4\sqrt{\pi }}ln\left[ d\left( 1-\frac{2c}{r}\right)
^{\frac{\alpha }{\sqrt{3}}}(qt+b)^{\sqrt{3}}\right] .  \label{psi}
\end{equation}%
Here, $q$, $b$, $c$ and $d$ are constants. Now under the transformation (\ref%
{conformal}), where $\phi =\exp [\psi \sqrt{\frac{8\pi }{\omega +\frac{3}{2}}%
}],$ we obtain

\begin{eqnarray}
d\bar{s}^{2}=\frac{B(t)^{1-\sqrt{3}/\beta }}{d^{1/\beta }A(r)^{\alpha /\sqrt{%
3}\beta }}\left[ A(r)^{\alpha }dt^{2}-A(r)^{-\alpha }dr^{2}\right] - \\ \frac{%
A(r)^{1-\alpha \frac{1+\sqrt{3}\beta }{\sqrt{3}\beta }}B(t)^{1-\sqrt{3}%
/\beta }r^{2}}{d^{1/\beta }}(d\theta ^{2}+sin^{2}\theta d\phi ^{2})
\nonumber \label{JBDsolution1}
\end{eqnarray}%
and

\begin{equation}
\phi (r,t)=\left[ dA(r)^{\alpha /\sqrt{3}}B(t)^{\sqrt{3}}\right] ^{1/\beta }
\label{phi1}
\end{equation}%
where $A(r)=1-\frac{2c}{r}$, $B(t)=qt+b$ and $\beta =\pm \sqrt{2\omega +3}$.
We now assume that $q\neq 0$ (i.e. the metric is not static) and define the
new time coordinate $\bar{t}=(qt+b)^{\frac{3}{2}-\frac{\sqrt{3}}{2\beta }}$.

In the limit that $c\rightarrow 0$ the $r$-dependence of the metric is
removed and the space becomes homogeneous. In this case we expect (\ref%
{JBDsolution1}) to reduce to the FRW Brans-Dicke metric given in the last
section. We see from the form of (\ref{JBDsolution1}) that the metric should
reduce to that of a flat FRW Brans-Dicke universe. Insisting on this limit
requires us to set $\beta =\sqrt{2\omega +3}$, $q=\frac{2\beta }{3\beta -%
\sqrt{3}}$ and $d=1$. This leaves the metric:

\begin{multline}
d\bar{s}^{2}=A(r)^{\alpha (1-\frac{1}{\sqrt{3}\beta })}d\bar{t}%
^{2}\\-A(r)^{-\alpha (1+\frac{1}{\sqrt{3}\beta })}\bar{t}^{\frac{2(\beta -%
\sqrt{3})}{3\beta -\sqrt{3}}}\times\\\left[ dr^{2}+A(r)r^{2}(d\theta ^{2}+\sin^{2}\theta
d\phi ^{2})\right] .  \label{JBDsolution2}
\end{multline}%
Rewriting (\ref{phi1}) with these coordinates and constants gives

\begin{equation}
\phi(r,t)=\left( 1-\frac{2c}{r} \right)^{\pm\frac{1}{2\beta}}\bar{t}^{2/(%
\sqrt{3}\beta -1)}.  \label{phi2}
\end{equation}

A comparison of (\ref{JBDsolution2}) with (\ref{phi_b(t)}) shows
that (\ref{JBDsolution2}) does indeed reduce to a flat vacuum FRW metric in
the limit $c\rightarrow 0$ (an inhomogeneous universe requires $c\neq 0$).
The metric (\ref{JBDsolution2}) is asymptotically flat and has singularities
at $\bar{t}=0$ and $r=2c$; the coordinates $r$ and $\bar{t}$ therefore cover
the ranges $0\leq \bar{t}<\infty $ and $2\leq \frac{r}{c}<\infty $.

\begin{figure}%[tbh]
\epsfig{file=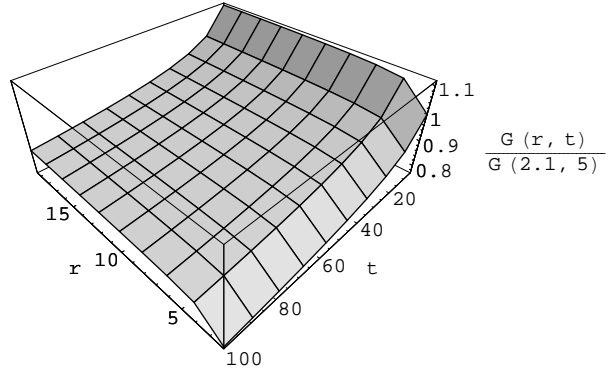,height=6.5cm}
\caption{{\protect {\textit{This graph illustrates the possible space and time
variations that can arise in G in inhomogeneous solutions to the Brans--Dicke
field equations, normalised at $r=2.1$ and $t=5$.}}}}
\label{GTB}
\end{figure}

The equations (\ref{Gc}) and (\ref{phi2}) can now be used to construct
a plot of $G(r,t)$; this is done in Figure \ref{GTB} which
was constructed by choosing $-\frac{1}{2\beta }$ in (\ref{phi2}). From the form of $%
G(r,t) $ we see that this choice corresponds to an overdensity in the
mass distribution (identified by comparison with the inhomogeneous Brans--Dicke
solution with matter, found below). In this figure, $\omega $ was set equal
to $100$, and $c$ set equal to $1$. This plot shows how $G$ can vary in
space and time in an inhomogeneous universe which consists of a static
Schwarzschild-like mass sitting at $r=0$ in an expanding universe. As $%
r\rightarrow 0$ the solution approaches the behaviour of the static
spherical vacuum BD solution but as $r\rightarrow \infty $ it approaches the
behaviour of a BD Friedmann universe.

\subsection{An Inhomogeneous Brans-Dicke Solution With Matter}

\label{vacsol}

\begin{figure}%[tbh]
\epsfig{file=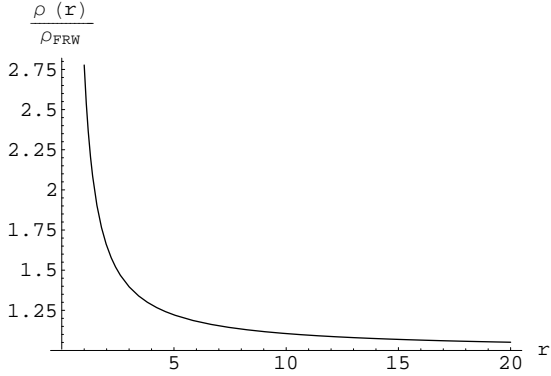,height=5cm}
\caption{{\protect {\textit{Distribution of $\protect\rho $ as a
function of $r$, with $\protect\omega =100$, from eq. (\protect\ref{rho(rt)}%
) and }}}${\protect {\mathit{c=0.5}}}$.}
\label{rho}
\end{figure}

We now seek a solution of the Brans-Dicke field equations (\ref{field
equations}) with the form

\begin{equation}  \label{metric}
ds^2=e^{\nu}dt^2-e^{\mu}a^2(dr^2+r^2d\theta^2+r^2 \sin^2\theta d\phi^2)
\end{equation}
where $e^{\nu }=e^{\nu (r)}$, $e^{\mu }=e^{\mu (r)}$ and $a=a(t)$. In
Appendix \ref{appendix} we show that a solution of the field equations (\ref%
{field equations}) for a metric of the form (\ref{metric}) is given by

\begin{equation}
e^{\nu }=\left( \frac{1-\frac{c}{2kr}}{1+\frac{c}{2kr}}\right) ^{2k},
\label{ev}
\end{equation}

\begin{equation}
e^{\mu }=\left( 1+\frac{c}{2kr}\right) ^{4}\left( \frac{1-\frac{c}{2kr}}{1+%
\frac{c}{2kr}}\right) ^{2(k-1)(k+2)/k},  \label{eu}
\end{equation}

\begin{equation}  \label{a}
a(t)=a_0\left(\frac{t}{t_0}\right)^{\frac{2\omega(2-\gamma)+2}{%
3\omega\gamma(2-\gamma)+4}},
\end{equation}
and

\begin{equation}
\phi (r,t)=\phi _{0}\left( \frac{t}{t_{0}}\right) ^{\frac{2(4-3\gamma )}{%
3\omega \gamma (2-\gamma )+4}}\left( \frac{1-\frac{c}{2kr}}{1+\frac{c}{2kr}}%
\right) ^{-2(k^{2}-1)/k}  \label{phi(rt)}
\end{equation}%
for the matter distribution

\begin{equation}
\rho (r,t)=\rho _{0}\left( \frac{a_{0}}{a(t)}\right) ^{3\gamma }\left( \frac{%
1-\frac{c}{2kr}}{1+\frac{c}{2kr}}\right) ^{-2k}  \label{rho(rt)}
\end{equation}%
where $k=\sqrt{\frac{4+2\omega }{3+2\omega }}$. This separable solution
displays the same time dependence as the power--law FRW Brans-Dicke
universes, (\ref{bds1})--(\ref{bds2}), but with an additional inhomogeneous $%
r$--dependence created by the matter source at $r=0$. Such a distribution of
matter in space is illustrated by figure \ref{rho}. Here we have chosen, for
illustrative purposes, $\gamma =1$, $\omega =100$ and a background value set
by the choice $\rho =\rho _{FRW}$ ($\rho _{FRW}$ being the matter density that
would be expected in the corresponding homogeneous universe). The temporal
evolution of $\rho $ is exactly the same as the FRW case.

We see from Figure \ref{rho} that the matter density is isotropic and
asymptotically constant as $r\rightarrow \infty $ with a sharp power-law
peak near the origin. Now (\ref{phi(rt)}) gives us

\begin{equation}
G(r,t)=G_{0}\left( \frac{1-\frac{c}{2kr}}{1+\frac{c}{2kr}}\right)
^{2(k^{2}-1)/k}t^{-\frac{2(4-3\gamma )}{3\omega \gamma (2-\gamma )+4}}.
\label{G(rt)}
\end{equation}

\begin{figure}%[tbh]
\epsfig{file=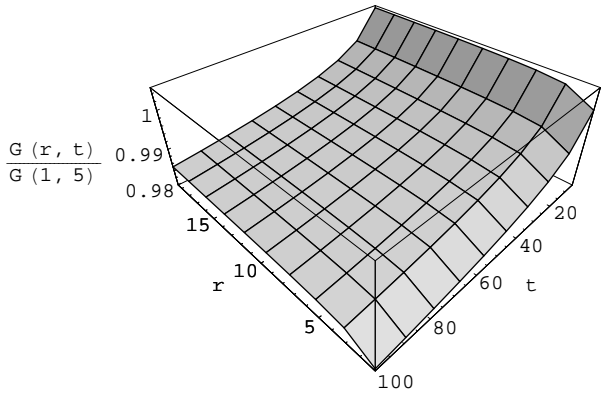,height=6.5cm}
\caption{{\protect {\textit{Evolution of $G(r,t)$ in space and time in
an inhomogeneous matter dominated Universe, with $\protect\omega =100$, from
eq. (\protect\ref{G(rt)}) and }}}$c=0.5.$}
\label{GTB3}
\end{figure}

Equation (\ref{G(rt)}) is used, with the values $\omega =100$, $\gamma =1$
and $c$ $=0.5$ to create Figure \ref{GTB3}, which shows the space--time
evolution of $G(r,t)$.

These results show how $G(r,t)$ can vary in space and time in an
asymptotically--flat universe with a peak of matter at the origin. Observers
located near the mass concentration will determine different values of $G$
locally although they will find the same values of $\dot{G}/G$ everywhere
because of the separable nature of the $G(r,t)$ evolution in eq. (\ref%
{rho(rt)}). This was also the case for solution (\ref{phi2}) given in
subsection \ref{vacsol}. In the next section we shall consider a more
realistic model in which both $G$ and $\dot{G}/G$ are different from place
to place. Plots like Figure \ref{GTB3} can be generated for universes
dominated by other types of cosmological fluid and with different rates of
density fall off with $r$.

\section{Matching Two Vacuum FRW Brans-Dicke Universes}

\label{vaccollapse}

We will now consider a simple model of a spherically symmetric cosmological
inhomogeneity produced by matching together flat and positively curved
vacuum FRW--BD universes. This is a well studied technique, first introduced
by Lema\^{\i}tre, for studying the non-linear evolution of overdensities in
general relativistic FRW universes. The overdense region is modelled as a
closed universe that at first expands more slowly than the background,
before reaching an expansion \ maximum and collapsing back to high density,
whilst the background continues to expand. In this section we consider
vacuum universes only, so there exists spherically symmetric inhomogeneity
in the expansion rate and in $\phi \sim G^{-1},$ but $\rho =p=0$.

For flat vacuum FRW universes eq. (\ref{Friedmann}) gives

\begin{equation}
\left( \frac{\dot{a}}{a}\right) ^{2}+\frac{\dot{a}}{a}\frac{\dot{\phi _{b}}}{%
\phi _{b}}=\frac{\omega }{6}\left( \frac{\dot{\phi _{b}}}{\phi _{b}}\right)
^{2}  \label{backFried}
\end{equation}%
where $\dot{}=\frac{d}{dt}$ and $\phi _{b}=\phi _{b}(t)$ is the BD scalar
field and $a$ is the scale factor in the flat background. For a positively curved ($k=+1$) region the
scale factor is taken to be $S(\tau)$, which satisfies the Friedmann
equation for the closed vacuum BD universe:

\begin{equation}
\left( \frac{S^{\prime }}{S}\right) ^{2}+\frac{S^{\prime }}{S}\frac{\phi
_{p}^{\prime }}{\phi _{p}}=\frac{\omega }{6}\left( \frac{\phi _{p}^{\prime }%
}{\phi _{p}}\right) ^{2}-\frac{k}{S^{2}},  \label{pertFried}
\end{equation}%
where $^{\prime }=\frac{d}{d\tau }$, $\phi _{p}=\phi _{p}(\tau )$ and $\tau $
and $k$ are the proper time and curvature of this perturbed region.

In matching these two regions at $t=t_0=\tau_0$ we must satisfy the boundary
conditions

\begin{align*}
S(\tau _{0})& =a(t_{0}),& \left( \frac{dS}{d\tau }\right) _{0}&=\left( 
\frac{da}{dt}\right) _{0},
\\
\phi_{p}(\tau_0)& =\phi _{b}(t_0)& \text{and} \quad
\left( \frac{d\phi _{p}}{d\tau }\right) _{0}& =\left( \frac{d\phi _{b}}{%
dt}\right) _{0}.
\end{align*}

\subsection{The Background Universe}

Assuming solutions of the form $\phi _{b}\propto t^{x}$ and $a\propto t^{y}$
and setting $a(0)=0$ gives, on substitution into (\ref{backFried}) and (\ref%
{Nariai2}), the $k=0$ BD vacuum solutions \citep{Tupper}

\begin{equation}
a(t)=t^{\frac{1}{3}(1+2(1-\sqrt{3(3+2\omega )})^{-1})}\label{phi_b(t)}
\end{equation}
and
\begin{equation}
\phi _{b}(t)=\phi _{b0}\left( \frac{t}{t_{0}}\right) ^{-2(1-\sqrt{%
3(3+2\omega )})^{-1}}. \nonumber 
\end{equation}

\subsection{A Collapsing Universe}

For the closed region we now follow the method given in refs. \citep{bar,
Barrow and Parsons} to find expressions for $S(\tau )$ and $\phi _{p}(\tau )$%
. We start by introducing conformal time, $\eta $, defined by $Sd\eta =d\tau 
$; then eq. (\ref{Nariai2}) becomes

\begin{equation*}
\phi _{p},_{\eta \eta }+\frac{2}{S}S,_{\eta }\phi _{p},_{\eta }=0.
\end{equation*}
This integrates directly to yield

\begin{equation}
\phi _{p},_{\eta }S^{2}=\sqrt{3}A(2\omega +3)^{-1/2}  \label{phi,n}
\end{equation}%
where $A$ is a constant. We now introduce the variable $y=\phi _{p}S^{2}$ to
write (\ref{Friedmann}) as

\begin{equation}
y,_\eta^2 = -4ky^2 + \frac{1}{3}\phi_p,_\eta^2 S^4 (2 \omega +3).
\label{y,n}
\end{equation}

Now, eqs. (\ref{phi,n}) and (\ref{y,n}) give

\begin{equation}
\frac{\phi_p,_\eta}{\phi_p} = \sqrt{3} A y^{-1} (2 \omega +3)^{-1/2} \nonumber
\end{equation}
and
\begin{equation}
y,_\eta^2 = -4ky^2 + A^2.  \label{y,n2}
\end{equation}

The solutions of eqs. (\ref{y,n2}), when $k > 0$, are given by

\begin{equation}
y(\eta) = \frac{A}{2\sqrt{k}} \sin(2\sqrt{k}(\eta+B),
\end{equation}
and

\begin{equation}
\phi _{p}(\eta )=C\tan ^{\sqrt{\frac{3}{(2\omega +3)}}}(\sqrt{k}(\eta +B))
\label{phi_p(n)}
\end{equation}%
where $B$ and $C$ are arbitrary constants. We now fix the conformal time
origin by setting $B=0$, so that $y=\phi _{p}S^{2}$ gives

\begin{equation}
S(\eta) \propto \frac{\sin^{1/2}(2\sqrt{k}\eta)}{\tan^{\sqrt{\frac{3}{%
4(2\omega +3)}}}(\sqrt{k}\eta)}.  \label{S(n)}
\end{equation}

\subsection{From $\protect\eta$ to $t$}

The function $\tau (\eta )$ is now obtained by integrating $Sd\eta =d\tau $
and $\tau (\eta )$ can then be used to obtain $S(\tau )$. We now require a
relation between $t$ and $\tau $, for this we proceed as in ref. \citep%
{Barrow and Kunze} and use the equation of relativistic hydrostatic
equilibrium \citep{Harrison7073,Harrison70731}, \citep{LandauLifschitz}

\begin{equation}
\frac{\partial \Phi }{\partial r}=-\frac{\partial p/\partial r}{p+\rho }
\label{hydroeq}
\end{equation}%
where $\Phi $ is the Newtonian gravitational potential and $r$ is radial
distance. Eqn. (\ref{hydroeq}) is derived under the assumptions that the
configuration is static and the gravitational field is weak, so $\Phi $
completely determines the metric; then (\ref{hydroeq}) is given by the
conservation of energy-momentum for a perfect fluid. We now use $d\tau
=e^{\Phi }dt$, and for a scalar field we have an effective state with $%
p=\rho $ and $\rho =\frac{\omega }{\phi }\dot{\phi}^{2}$. Combining these
results gives

\begin{equation}
\frac{d\tau}{dt} = \frac {\dot{\phi_b}}{\phi_p^{\prime}} \frac{\phi_p^{1/2}}{%
\phi_b^{1/2}}.  \label{dT/dt}
\end{equation}

Now $\dot{\phi}_{b}\propto a^{-3}$ and $\phi _{p}^{\prime }\propto S^{-3}$,
and so with (\ref{phi_p(n)}) and (\ref{phi_b(t)}) this gives

\begin{equation}
\frac{d\tau }{dt}=\frac{\sin ^{1/2}(2\sqrt{k}\eta )}{\sin ^{1/2}(2\sqrt{k}%
\eta _{0})}\frac{S^{2}}{S_{0}^{2}}\frac{a^{\sqrt{1+\frac{2}{3}\omega }-4}}{%
a_{0}^{\sqrt{1+\frac{2}{3}\omega }-4}}  \label{dTdt2}
\end{equation}

Eq. (\ref{dTdt2}) and the relation $Sd\eta =d\tau $ allow us to obtain $S(t)$
from (\ref{S(n)}). This is done numerically. Now fixing the constants of
proportionality together with $k$ in eqs. (\ref{S(n)}), (\ref{phi_p(n)}) and
(\ref{phi_b(t)}), in order to satisfy the boundary conditions, we find
equations for the evolution of the scale factors and scalar fields in the
flat background and perturbed region. These are matched at a boundary, at
time $t_{0}=\tau _{0}=\eta _{0}$.

\subsection{Results}

\begin{center}
\begin{figure}%[htb]
\epsfig{file=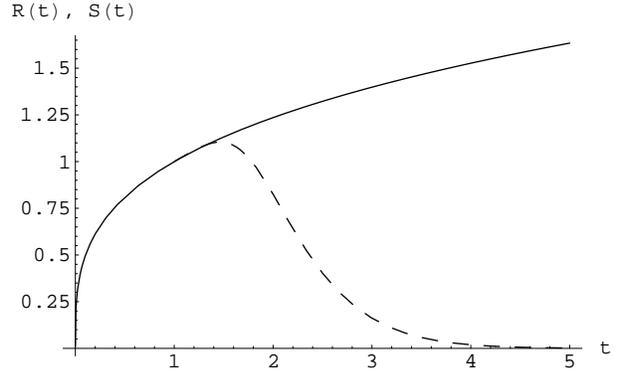,height=5cm}
\caption{{\protect {\textit{Evolution of the scale factor $S$ in the
perturbed overdense region (dashed line) and in the background $a$(solid
line) with respect to the comoving proper time in the flat background.}}}}
\label{RandS}
\end{figure}
\end{center}

Figure \ref{RandS} shows the evolution of $a(t)$ and $S(t)$ when the region
described by $S(t)$ becomes positively curved at initial time $t_{0}=1$. In
Figure \ref{RandS} we choose $\omega =100$, for illustrative purposes, and $%
a_{0}=S_{0}=1$ so that the boundary condition for the matching of the first
derivatives of the scale factors is given by $\left( \frac{dS}{d\tau }%
\right) _{0}=\left( \frac{dS}{d\eta }\right) _{0}=\left( \frac{da}{dt}%
\right) _{0}.$

\begin{figure}%[tbh]
\epsfig{file=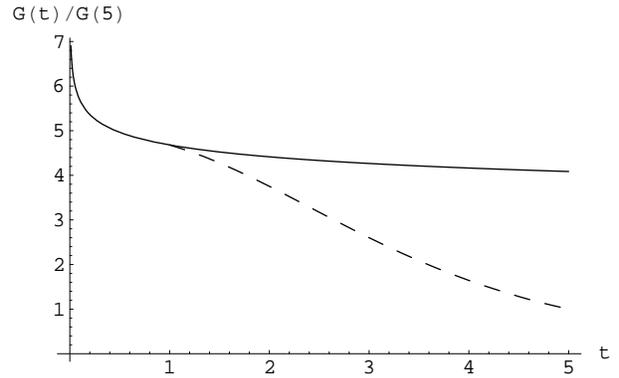,height=5cm}
\caption{{\protect {\textit{Evolution of $G(t)$ in the overdense
perturbed overdense region of positive curvature (dashed line) and in the
spatially flat background universe (solid line).}}}}
\label{FRWG}
\end{figure}

We can now express $G=G(t)$ in the regions of different curvature using
equations (\ref{Gc}), (\ref{phi_p(n)}) and (\ref{phi_b(t)}), along with the
appropriate coordinate transformations. This gives Figure \ref{FRWG}, below.
It is clearly seen that the evolution of $G(t)$ is quite different in the
two regions, as expected. The collapsing overdensity evolves faster than the
background and possesses a smaller value of $G$ but a larger value of $|\dot{%
G}/G|\ $at all times after the evolution commences. 

\section{A More Refined Spherical Collapse Model}

\label{Results}

We now generalise the spherical collapse model described in the last section
to the more astronomically realistic case of a flat universe containing
matter and a cosmological constant (see. e.g. refs. \citep{Padmanabhan}, \citep%
{peacock} or \citep{lahav}). As before, we match a flat Brans-Dicke FRW
background to a spherically symmetric overdensity at an appropriate boundary
and allow the two regions to evolve separately.

\subsection{The background universe}

\label{background}

Again, we consider a flat $(k=0)$, homogeneous and isotropic background
universe. Since we are interested in the matter-dominated epoch, when
structure formation starts, we can assume that our universe contains only
matter and a vacuum energy contribution so that $\rho =\rho _{m}+\rho
_{\Lambda }$ and $p=p_{\Lambda }=-\rho _{\Lambda }$ give the total density
and pressure, respectively. So, for the flat background, eq. (\ref{Friedmann}%
) gives for a general $\omega (\phi )$ theory,

\begin{equation}
\left( \frac{\dot{a}}{a}\right) ^{2}+\frac{\dot{a}}{a}\frac{\dot{\phi}}{\phi 
}=\frac{\omega }{6}\frac{\dot{\phi}^{2}}{\phi ^{2}}+\frac{8\pi }{3\phi }%
(\rho _{m}+\rho _{\Lambda })  \label{fried}
\end{equation}%
and eq. (\ref{Nariai2}) gives

\begin{equation}
\ddot{\phi}+3\frac{\dot{a}}{a}\dot{\phi}=\frac{8\pi }{(2\omega +3)}(\rho
_{m}+4\rho _{\Lambda })-\frac{\dot{\omega}\dot{\phi}}{(2\omega +3)}.
\label{phidot}
\end{equation}

Here, $\rho _{m}\propto a(t)^{-3}$ and $\rho _{\Lambda }=\text{constant}$.
These equations govern the evolution of $\phi (t)$ and $a(t)$ in the flat
expanding cosmological background. The contribution of the vacuum energy
stress ($p=-\rho $) to the Friedmann equation (\ref{fried}) in BD cosmology
differs from that in general relativity because of the presence of the
variable $\phi $ field: it is not the same as the addition of a cosmological
constant term to the RHS of (\ref{fried}). However, with this proviso, we
shall continue to refer to $\Lambda $cdm models in Brans-Dicke theories in
the following sections.

\subsection{The overdensity}

\label{perturbation}

Again, we consider a spherical overdense region of radius $S$ and model the
interior space-time as a closed FRW Brans-Dicke theory universe, ignoring
any anisotropic effects of gravitational instability or collapse. As usual,
we assume there is no shell-crossing; this implies mass conservation inside
the overdensity and independence of the radial coordinate \citep{Padmanabhan}%
. The evolution equations can now be written in a form that ignores the
spatial dependence of the fields. Put $\rho =\rho _{cdm}+\rho _{\Lambda }$
and $p=p_{\Lambda }=-\rho _{\Lambda }$, where 'cdm' corresponds to cold dark
matter, so that (\ref{acceleration}) gives

\begin{multline}
\ddot{S}-\dot{S}\frac{\dot{\phi _{c}}}{\phi _{c}}=-S\left( \frac{\omega _{c}
}{3}\frac{\dot{\phi _{c}}^{2}}{\phi _{c}^{2}}-\frac{1}{2}\frac{\dot{\omega}_{c}}{(2\omega _{c}+3)}\frac{\dot{\phi}
_{c}}{\phi _{c}}  \right.\\\left.  + \frac{8\pi }{3\phi _{c}}\frac{
(\rho _{cdm}(3+\omega _{c})+\rho _{\Lambda }(3-2\omega _{c}))}{(2\omega
_{c}+3)} \right) \label{rc}
\end{multline}%
while (\ref{Nariai2}) reduces to

\begin{equation}
\ddot{\phi _{c}}+3\frac{\dot{S}}{S}\dot{\phi _{c}}=\frac{8\pi }{(2\omega
_{c}+3)}(\rho _{cdm}+4\rho _{\Lambda })-\frac{\dot{\omega _{c}}\dot{\phi _{c}%
}}{(2\omega _{c}+3)}  \label{phic}
\end{equation}%
where $S=S(t)$ is the scale factor and $\phi _{c}=\phi _{c}(t)$ is the BD
scalar field in the collapsing region of positive curvature where $\rho
_{cdm}\propto S(t)^{-3}$ and $\rho _{\Lambda }=\text{constant}$. These
equations give the evolution of $\phi _{c}(t)$ and $S(t)$.

We have assumed that the equation of motion of the field inside the cluster
overdensity is described by the local space-time geometry. This means that
the field follows the dark-matter collapse from the beginning of the
cluster's formation. We do not consider this to be fully realistic since
there is expected to be an outflow of energy associated with $\phi $ from
the overdensity to the background universe, as first noticed by Mota
and van de Bruck (\cite{carsten}).  The details of this outflow of
energy and its effect on the collapse can only be determined by a fully
relativistic hydrodynamical calculation, which is beyond the scope of this
study. Nevertheless, at late times during the collapse of the dark matter
(and especially when the density contrast in the dark matter is very large)
the field should no longer feel the effects of the expanding background and
will decouple from it. We are also neglecting the effects of deviations from
spherical symmetry, which grow during the collapse in the absence of
pressure, along with rotation, gravitational tidal interactions between
different overdensities, and all forms of non-linear hydrodynamical
complexity.

\subsection{Evolution of the overdensity}

Consider the spherical perturbation in the cdm fluid with a spatially
constant internal density. Initially, this perturbation is assumed to have a
density amplitude $\delta _{i}>0$ where $|\delta _{i}|\ll 1$. The initial
cdm density inside the overdensity is therefore $\rho _{cdm}=\rho
_{m}(1+\delta _{i})$.

Four characteristic phases of the overdensity's evolution can be identified:

\begin{itemize}
\item \textit{Expansion}: we employ the initial boundary condition $\phi
_{c}=\phi $ and assume that at early times the overdensity expands along
with the background.

\item \textit{Turnaround}: for a sufficiently large $\delta _{i},$ gravity
prevents the overdensity from expanding forever; the spherical overdensity
breaks away from the general expansion and reaches a maximum radius.
Turnaround is defined as the time when $S=S_{max}$, $\dot{S}=0$ and $\ddot{S}%
<0$.

\item \textit{Collapse}: the overdensity subsequently collapses ($\dot{S}<0$%
). If pressure and dissipative physics are ignored the overdensity would
collapse to a singularity where the density of matter would tend to
infinity. In reality this singularity does not occur; instead, the kinetic
energy of collapse is transformed into random motions.

\item \textit{Virialisation}: dynamical equilibrium is reached and the
system becomes stationary with a fixed radius and constant energy density.
\end{itemize}

We require our spherical overdensity to evolve from the linear perturbation
regime at high redshift until it becomes non-linear, collapses, and
virialises. Thereafter, the overdensity will become gravitationally stable
and further local evolution of the scale factor and of the scalar field will
cease. However, the background scale factor will continue to expand, and so
the background density and background value of $G$ will continue to
decrease. As a result, a significant disparity between the evolution of $G$
inside and outside the cluster can result.

\subsection{Virialisation}

In scalar-tensor theories we expect that the gravitational potential will
not be of the standard local $r^{-1}$ form. This requires reconsideration of
the virial condition. According to the virial theorem, equilibrium will be
reached when \citep{Goldstein}

\begin{equation}
T=\frac{1}{2}S\frac{\partial U}{\partial S}  \label{virial}
\end{equation}%
where $T$ is the average total kinetic energy, $U$ is the average total
potential energy and $S$ here denotes the radius of the spherical
overdensity.

The potential energy for a given component $x$ can be calculated from its
general form in a spherical region \citep{LandauLifschitz}

\begin{equation*}
U_{x}=2\pi \int_{0}^{S}\rho _{tot}\Phi _{x}r^{2}dr,
\end{equation*}%
where $\rho _{tot}$ is the total energy density and $\Phi _{x}$ is the
gravitational potential due to the density component $\rho _{x}$.

The gravitational potential $\Phi _{x}$ can be obtained from the weak-field
limit of the field equations (\ref{field equations}). This results in a
Poisson equation where the terms associated to the scalar field can be
absorbed into the definition of the Newtonian constant as \citep{Will}

\begin{equation}  \label{Gc}
G_c=\frac{4+2\omega_c(\phi_c)}{3+2\omega_c(\phi_c)}\frac{1}{\phi_c}.
\end{equation}

This results in the usual form for the Newtonian potential

\begin{equation*}
\Phi _{x}\left( s\right) =-2\pi G_{c}\rho _{x}(3\gamma _{x}-2)\left( S^{2}-%
\frac{r^{2}}{3}\right)
\end{equation*}%
where $G_{c}$ is given by equation (\ref{Gc}) and $\gamma _{x}-1$ is $%
p_x/\rho_x$ for the fluid component with density $\rho _{x}$ and pressure $%
p_x$ (appearing due to the relativistic correction to Poisson's equation: $%
\Delta \Phi =4\pi G\left( \rho +3p\right) $).

In $\Lambda $cdm models of structure formation it is entirely plausible to
set $\gamma _{x}=1$ as the energy density of the cosmological constant is
negligible on the virialised scales we are considering \citep{wang,carsten}. The
potential energies associated with a given component ($x$) inside the
overdensity are now given by

\begin{equation}
U_{x}=-\frac{16\pi ^{2}}{15}G_{c}\rho _{tot}\rho _{x}S^{5}.
\end{equation}
Therefore, the virial theorem will be satisfied when

\begin{equation*}
T_{vir}=\frac{1}{2}S_{vir}\left( \frac{\partial U}{\partial S}\right) _{vir},
\end{equation*}%
where

\begin{multline}
\frac{\partial U}{\partial S}=-\frac{16\pi ^{2}}{15}\left[ \frac{\partial
G_{c}}{\partial S}\rho _{tot}\rho _{x}S^{5}+ \right.\\ \left.G_{c}\frac{\partial \rho _{tot}}{%
\partial S}\rho _{x}S^{5}+G_{c}\rho _{tot}\frac{\partial \rho _{x}}{\partial
S}S^{5}+G_{c}\rho _{tot}\rho _{x}5S^{4}\right] \label{du/dr}
\end{multline}
and we have used $U_{tot}=U_{cdm}+U_{\Lambda }+U_{\phi _{c}}$, $\rho
_{tot}=\rho _{cdm}+\rho _{\Lambda }+\rho _{\phi _{c}}$, $\rho _{\phi
_{c}}=\omega _{c}\dot{\phi}^{2}/\phi $ and

\begin{equation*}
\frac{\partial G_{c}}{\partial S}=\frac{\dot{G}_{c}}{\dot{S}}=-G_{c}\ \dot{%
\phi}_{c}\frac{3+2\omega _{c}}{4+2\omega _{c}}\left( G_{c}+\frac{2\omega
_{c}^{^{\prime }}(\phi _{c})}{(3+2\omega _{c})^{2}}\right) .
\end{equation*}%
The other components of eq. (\ref{du/dr}) are obtained from eqs. (\ref{rc})
and (\ref{phic}).

Using eq. (\ref{virial}), together with energy conservation at turnaround
and virialisation, we obtain an equilibrium condition in terms of potential
energies only

\begin{equation}
\frac{1}{2}S_{vir}\left( \frac{\partial U}{\partial S}\right)
_{z_{v}}+U_{tot}(z_{v})=U_{tot}(z_{ta}),  \label{virialcond}
\end{equation}%
where $z_{v}$ is the redshift at virialisation and $z_{ta}$ is the redshift
of the over-density at its turnaround radius. The behaviour of $G$ during
the evolution of an overdensity can now be obtained by numerically evolving
the background eq. (\ref{fried})-(\ref{phidot}) and the overdensity eqs. (%
\ref{rc}) and (\ref{phic}) until the virial condition (\ref{virialcond})
holds.

We point out here an inconsistency when one makes use of equation (\ref%
{virialcond}) together with the assumption that energy is not conserved.
This inconsistency is removed by assuming a negligible outflow of $\phi $
from the overdensity, in which case we regain energy conservation within the
system and so retain self-consistency.

\subsection{Overdensities vs Background}

\subsubsection{Brans--Dicke theory}
\label{bransd}

\begin{figure*}%[tbh]
\begin{center}
%$%
%\begin{array}{c@{\hspace{0.4in}}c}
%\multicolumn{1}{l}{\mbox{}} & \multicolumn{1}{l}{\mbox{}} \\[-0.53cm] 
%\epsfxsize=3in \epsffile{GcGbw40000.eps} & \epsfxsize=3in %
%\epsffile{GcGbw500.eps} \\[0.4cm] 
%\mbox{\textbf{(a)} $\omega=40000$} & \mbox{\textbf{(b)} $\omega=500$}%
%\end{array}%
%$%
\mbox{\subfigure[$\omega=40000$]{\epsfig{figure=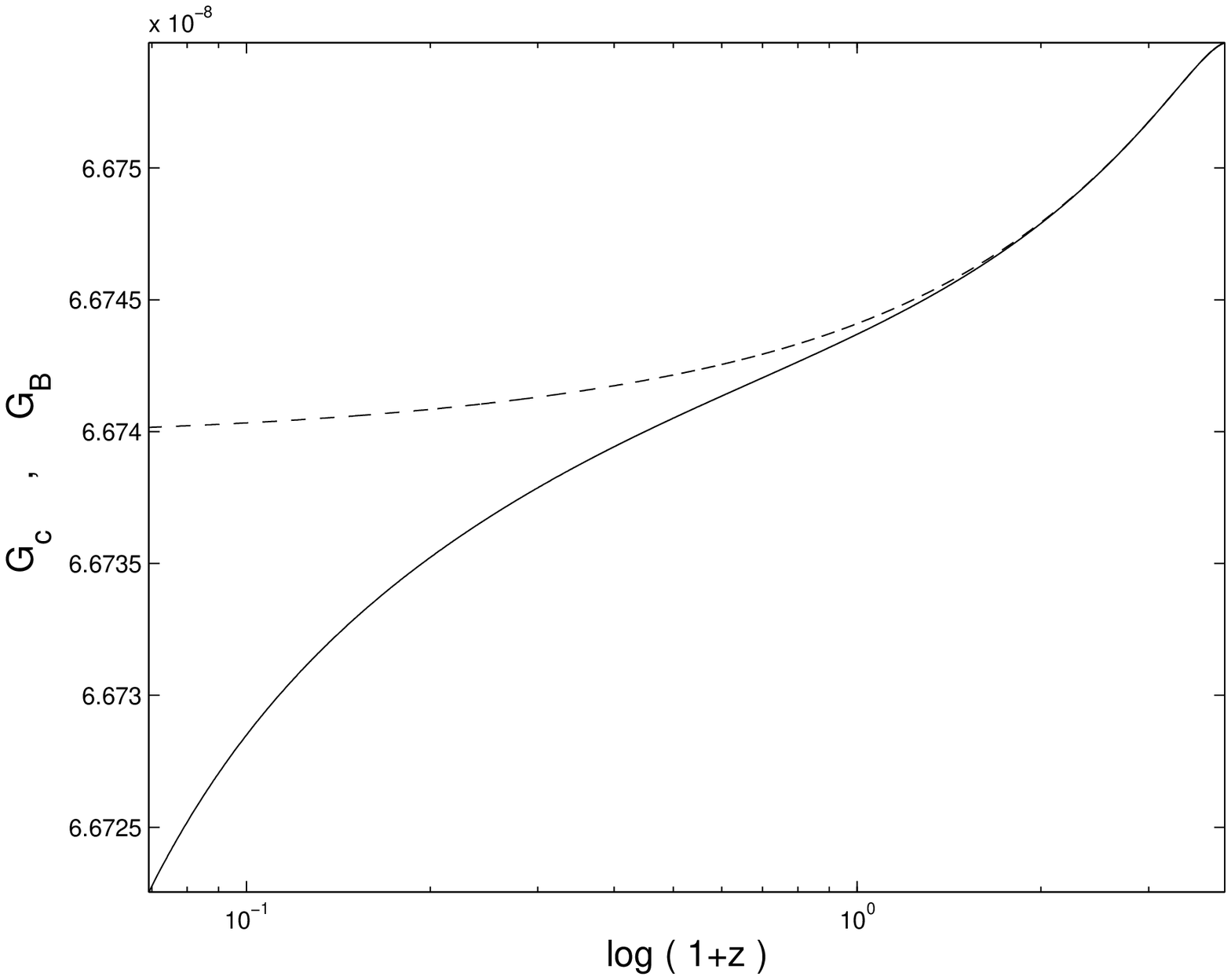, height=6cm}}\quad
\subfigure[$\omega=500$]{\epsfig{figure=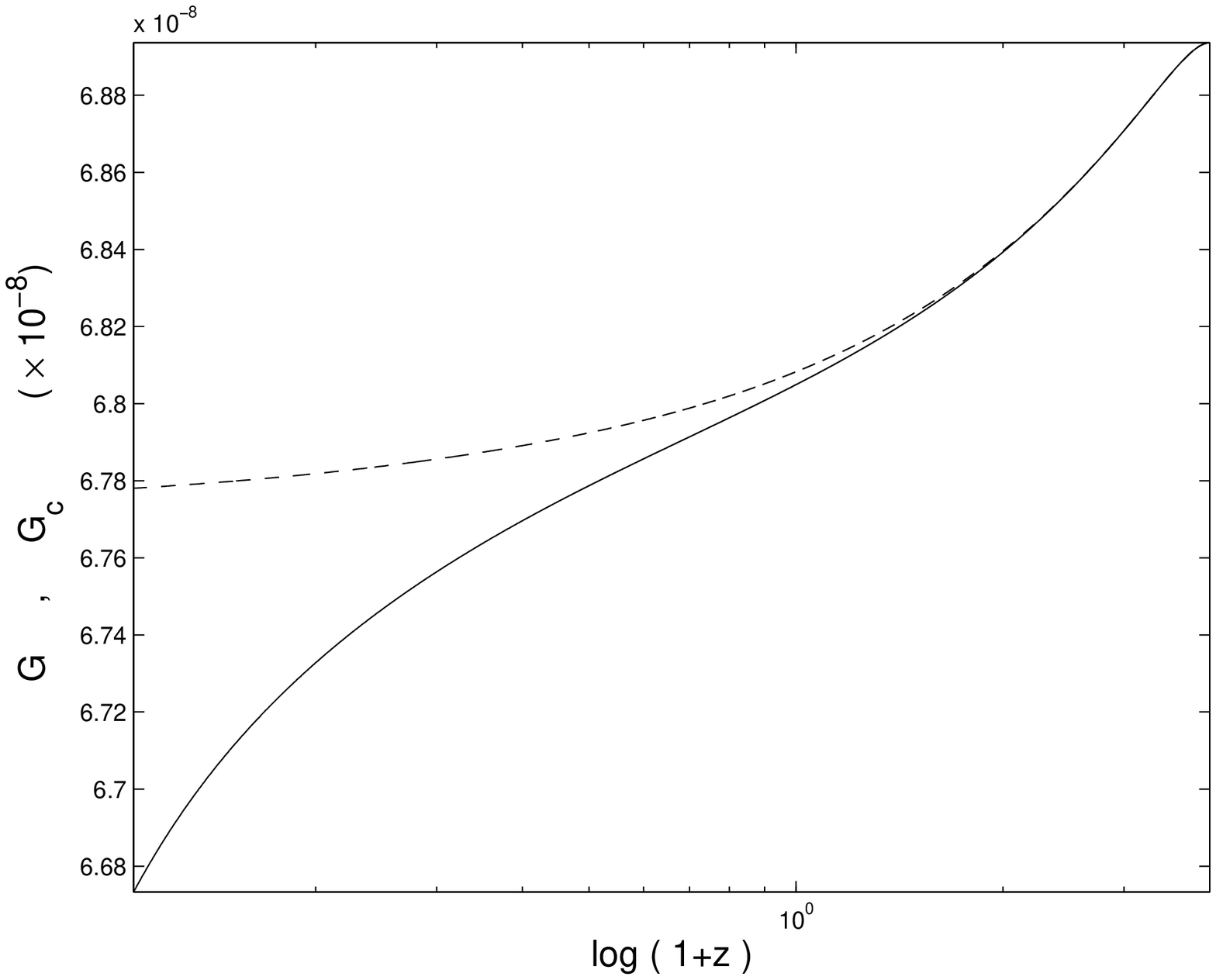, height=6cm}}}
\end{center}
\caption{{\protect {\textit{Plots of $G$ against $\ln (1+z)$ in the
background (dashed-line) and in an overdensity (solid-line), for different
values of $\protect\omega .$ Initial conditions are chosen in both cases so
as to give $G=G_{0}$, the present value of the Newton constant, at
virialisation.  We note that increasing $\omega$ decreases the
difference in $G$ between the overdensity and the background.}}}}
\label{JBDplot}
\end{figure*}

The Brans--Dicke coupling parameter $\omega $ is constant and constrained by
a variety of local gravitational tests (see \citep{Uzan} for a review). The
strongest constraint to date is derived from observations of the Shapiro
time delay of signals from the Cassini space craft as it passes behind the
Sun. These considerations led Bertotti, Iess and Tortora \citep{Bertotti},
after a complicated data analysis process, to claim that $\omega $ must have
a value greater than $40000$ (to $2\sigma $). This limit on $\omega $ must
be satisfied at all times in all parts of the universe, and leads to the
conclusion that Brans-Dicke theory must be phenomenologically very similar
to general relativity throughout most of the history of the universe.
However, we do still expect a cosmological evolution of the Brans-Dicke
field $\phi $ which determines the value of Newton's $G$; and we expect this
evolution to be different in regions that collapse to form the structure probed
by Cassini compared to that in the idealised expanding cosmological
background, as described above. Hence we expect the measurable value of $G$
to be different in these two distinct regions with different histories. It
is quite possible that the region of gravitational equilibrium probed by
Cassini and other local observations would find no perceptible change in $G$
locally despite the presence of change on cosmological scales outside of
bound inhomogeneities.

In this section we quantify this difference in $G$ by numerically evolving $%
a $, $S$, $\phi $ and $\phi _{c}$ until virialisation occurs. At
virialisation, the evolution of $S$ and $\phi _{c}$ is expected to end,
giving a value of $G$ that is locally constant in time even though the
cosmological evolution of $a$ and $\phi $ continues.

The plots in figure \ref{JBDplot} were constructed using the representative
values $\omega =40000$ and $\omega =500$ and the boundary condition $\phi
_{c0}\simeq G_{0}^{-1}$, so that the value of $G$ measured inside the
overdensity at present is equal to the value of Newton's $G$, as measured
locally. The evolution of the background was determined by matching $\phi
_{c}$ to $\phi $ at the time when the overdensity decouples from the
background, $t_{i}$. We see a clear difference in the evolution of $G$ in
the two regions, as expected.  This example shows that we expect different values of $G$ and $\dot{G}/G$
inside and outside virialised overdensities. The present value of $G$ and $%
\dot{G}/G$ depends on the history of the region where it was sampled,
as well as on the Brans--Dicke coupling parameter, $\omega$.  

It can
be seen from the plots in figure \ref{JBDplot} that increasing
$\omega$ has the effect of decreasing the difference in $G$
between the background universe and the overdensity.  The size of this
inhomogeneity is found to be of order $1/\omega$ and, correspondingly,
reduces to zero as $\omega \rightarrow \infty$.  This is an important
consistency check for the methods used as we expect Brans--Dicke
theory to reduce to general relativity, with a constant $G$, in this limit.

\subsubsection{Scalar-tensor theory with $2\protect\omega +3=2A\left\vert 1-%
\frac{\protect\phi }{\protect\phi _{\infty }}\right\vert ^{-p}$}
\label{theory1}

\begin{figure*}
\begin{center}
%$%
%\begin{array}{c@{\hspace{0.4in}}c}
%\multicolumn{1}{l}{\mbox{}} & \multicolumn{1}{l}{\mbox{}} \\[-0.53cm] 
%\epsfxsize=3in \epsffile{wgraph1.eps} & \epsfxsize=3in \epsffile{wgraph2.eps}
%\\[0.4cm] 
%\mbox{\textbf{(a)}} & \mbox{\textbf{(b)}}%
%\end{array}%
%$%
\mbox{\subfigure[$2\protect\omega (\protect\phi )+3$ for p=2 and A=1, 2
and 5 (solid, dashed and dotted lines, respectively).]{\epsfig{figure=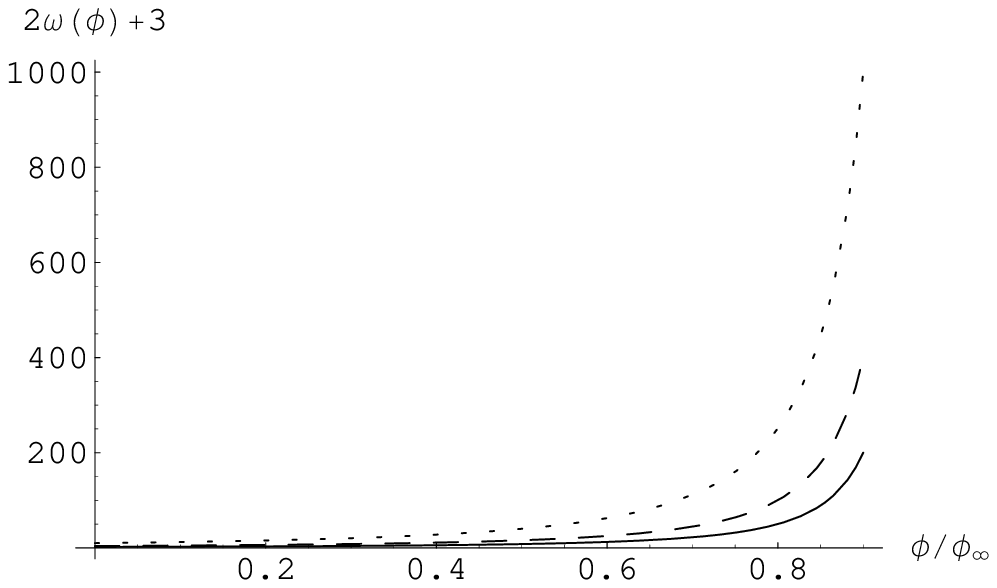, height=5cm}}\quad
\subfigure[ $2\protect%
\omega (\protect\phi )+3$ for A=1 and p=1.5, 2 and 2.5 (solid, dashed and
dotted lines, respectively).]{\epsfig{figure=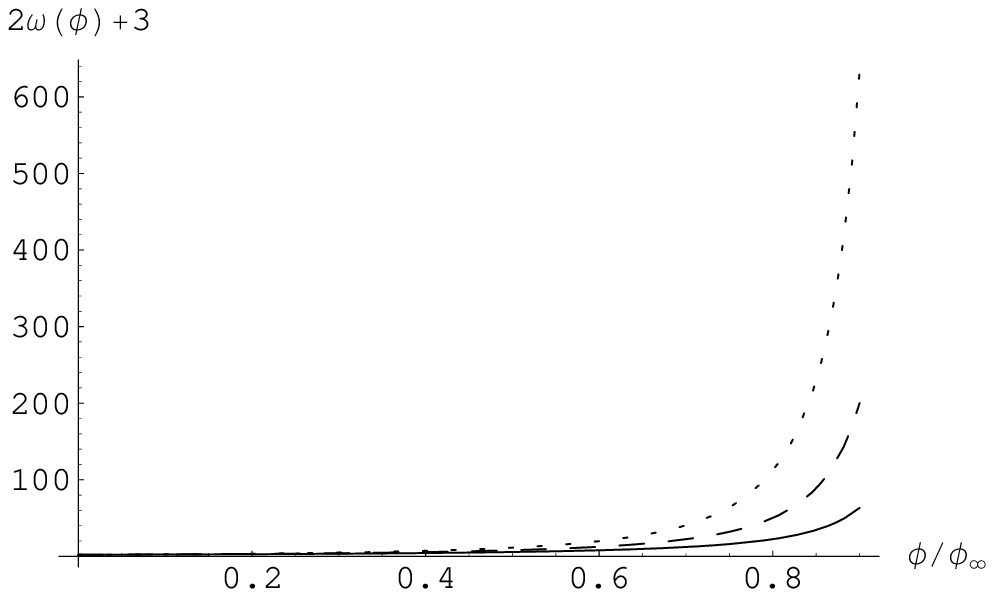, height=5cm}}}
\end{center}
%\par
\caption{
%\textbf{(a)} $2\protect\omega (\protect\phi )+3$ for p=2 and A=1, 2
%and 5 (solid, dashed and dotted lines, respectively). \textbf{(b)} $2\protect%
%\omega (\protect\phi )+3$ for A=1 and p=1.5, 2 and 2.5 (solid, dashed and
%dotted lines, respectively). 
In these gravity theories there is fast
approach to general relativity at late times when $\protect\phi \rightarrow 
\protect\phi _{\infty }$, but significantly different behaviour at early
times.}
\label{w(phi)}
\end{figure*}

Next, we consider a scalar--tensor theory with a variable $\omega (\phi )$.
We investigate the class of theories defined by the choice of coupling
function 
\begin{equation*}
2\omega (\phi )+3=2A\left\vert 1-\frac{\phi }{\phi _{\infty }}\right\vert
^{-p},
\end{equation*}
where $A$, $\phi _{\infty }$, and $p$ are positive definite constants. We
refer to this as Theory 1. Such a choice of coupling was considered by
Barrow and Parsons and was solved exactly for the case of a flat FRW
universe containing a perfect fluid \citep{Barrow and Parsons}.

Setting the constants as $2A=\left( \phi _{\infty }/\beta \right) ^{2}$ and $%
p=2$ gives us the scalar--tensor theory considered by Damour and Pichon \citep%
{DamourPichon} and by Santiago, Kalligas and Wagoner \citep{Santiago}. This
choice of $\omega (\phi )$ corresponds to setting $\ln A(\phi )=\ln A(\phi
_{0})+\frac{1}{2}\beta (\phi _{\infty }-\phi )^{2}$, where $A^{2}(\phi )$ is
the conformal factor $1/\phi$ from eq. (\ref{conformal}). Damour and Nordvedt \citep%
{DamourNordvedt} consider this function as a potential and therefore justify
its choice in relation to the generic parabolic form near a potential
minimum. Expecting the function to be close to zero (i.e. GR), Santiago,
Kalligas and Wagoner justify its expression as a perturbative expansion.
This choice of $\omega (\phi )$ with $p>1/2$ corresponds to a general
two--parameter class of scalar--tensor theories that are close to GR and
will be drawn ever closer to it with $\omega \rightarrow \infty $ and $%
\omega ^{\prime }/\omega ^{3}\rightarrow 0$ as the universe expands and $%
\phi \rightarrow \phi _{\infty }$. We therefore consider it as a
representative example of a wide family of plausible varying-$G$ theories
that generalise Brans-Dicke.

The evolution of this form of $\omega (\phi )$ is shown graphically in
figure \ref{w(phi)} for different values of $A$ and $p$. Clearly the
evolution of $\omega (\phi )$ is sensitive to both $A$ and $p$ and so the
choice of these parameters is important for the form of the underlying
theory. For illustrative purposes we choose here the values $p=1.5,2$
and $5$ and $A=1,2$ and $5$.

\begin{figure*}
\epsfig{file=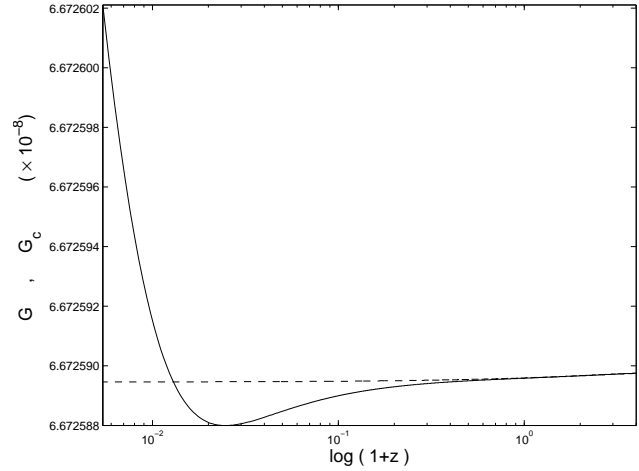,height=6.5cm}
\caption{{\protect {\textit{{$2\protect\omega +3=2A\left\vert 1-\frac{%
\protect\phi }{\protect\phi _{\infty }}\right\vert ^{-2}$ A graph of $\ $the
variation $G$ against $\ln (1+z)$ for the background universe (dashed-line)
and inside an overdensity (solid-line) which give $G=G_{0},$the presently
observed terrestrial value, at virialisation.}}}}}
\label{w(phi)plot}
\end{figure*}

In a similar way to the BD case we now create an evolution of $\phi_c$
that virialises at $z=0$ to give the value $G_{c0}=6.673\times
10^{-11}$, as observed experimentally.  The corresponding evolution
for $\phi_b$ is calculated as before by matching it to the value of
$\phi_c$ at the time the overdensity decouples from the background and
begins to collapse.  In creating these plots we have used the
conservative parameter values $p=2$, $\omega_{c0} = 1.2 \times10^5$ and 
$A= 6\times10^{-7}$ which are consistent with observation and allow
structure formation to occur in a similar way to general relativity.
The results of this are plotted in figure \ref{w(phi)plot}.

Again, we note the different evolution of $G(t)$ in the two regions, and the
difference in the asymptotic values of $G$. We note
that experimental measurements of $G$ on Earth have a significant
uncertainty with the 1998 CODATA value carrying an uncertainty 12 times 
\textit{greater} than the standard value adopted in 1987. The 1998 value is
given as \citep{codata, sch}

\begin{equation*}
G_{1998}=6.673\pm 0.010\times 10^{-8}cm^{3}gm^{-1}s^{-2},
\end{equation*}%
while the 2002 CODATA pre-publication announcement reverts to the earlier
higher accuracy consensus with \citep{codata2002}

\begin{equation*}
G_{2002}=6.6742\pm 0.0010\times 10^{-8}cm^{3}gm^{-1}s^{-2}\ 
\end{equation*}

We could re--run the above analysis with different values of $A$ and $p$,
but expect that the results would look qualitatively similar. From Figure %
\ref{w(phi)} we see that increasing (decreasing) the values of $A$ and $p$
will increase (decrease) the value of $\omega (\phi )$ for a given $\phi $,
thereby making the theory more (less) like GR. We therefore expect an
analysis with a higher (lower) value of $A$ and/or $p$ to look very similar
to the analysis presented above with a less (more) rapid evolution of $G(t)$%
. For the sake of brevity we omit such an analysis here. \emph{\ }

\subsection{Space and Time variations of $G$}

We now calculate how time and space variations of $G$ evolve with redshift
and depend on the cdm density contrast, $\Delta _{c}$. In order to do this,
we make the definitions

\begin{equation*}
\frac{\Delta G}{G}(t)\equiv \frac{G(t)-G_{0}}{G_{0}},\qquad \frac{\delta G}{G%
}(t)\equiv \frac{G_{c}(t)-G_{b}(t)}{G_{b}(t)}
\end{equation*}
and
\begin{equation*}
\Delta
_{c}\equiv \frac{\rho _{cdm}(z_{v})}{\rho _{m}(z_{v})},
\end{equation*}%
where $G_{c}$ and $G_{b}$ correspond to $G$ as measured in the overdensity
and in the background universe respectively.

\begin{figure}%[tbh]
\epsfig{file=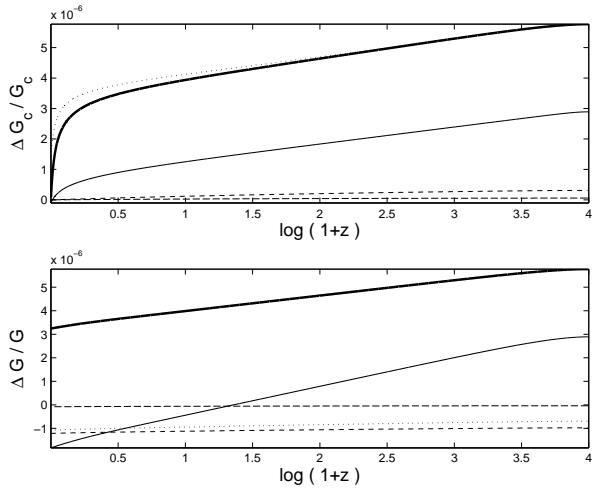,height=6.5cm}
\caption{{\protect {\textit{Evolution of $\Delta G/G$ as a function of 
$log(1+z)$ for overdensities which virialise at $z=0$ in a $\Lambda $cdm
model. Upper plot: evolution inside the overdensity. Lower plot: evolution
in the background universe. Thick solid line $2\protect\omega +3=4
\times 10^6$,
thin solid line $2\protect\omega +3=B^{2}\protect\phi $,
dashed-line $2\protect\omega +3=2A|1-\frac{\protect\phi }{\protect\phi _{0}}%
|^{-2}$, dash-dotted line $2\protect\omega +3=C|\ln (\frac{\protect\phi }{%
\protect\phi _{0}})|^{-4}$, dotted line $2\protect\omega +3=D|1-(\frac{%
\protect\phi }{\protect\phi _{0}})^{2}|^{-1}$ . Each model is normalised in
order to have $G_{0}=G_{c}(z=0)$ inside the overdensities.}}}}
\label{DeltaG}
\end{figure}

\begin{figure}%[tbh]
\epsfig{file=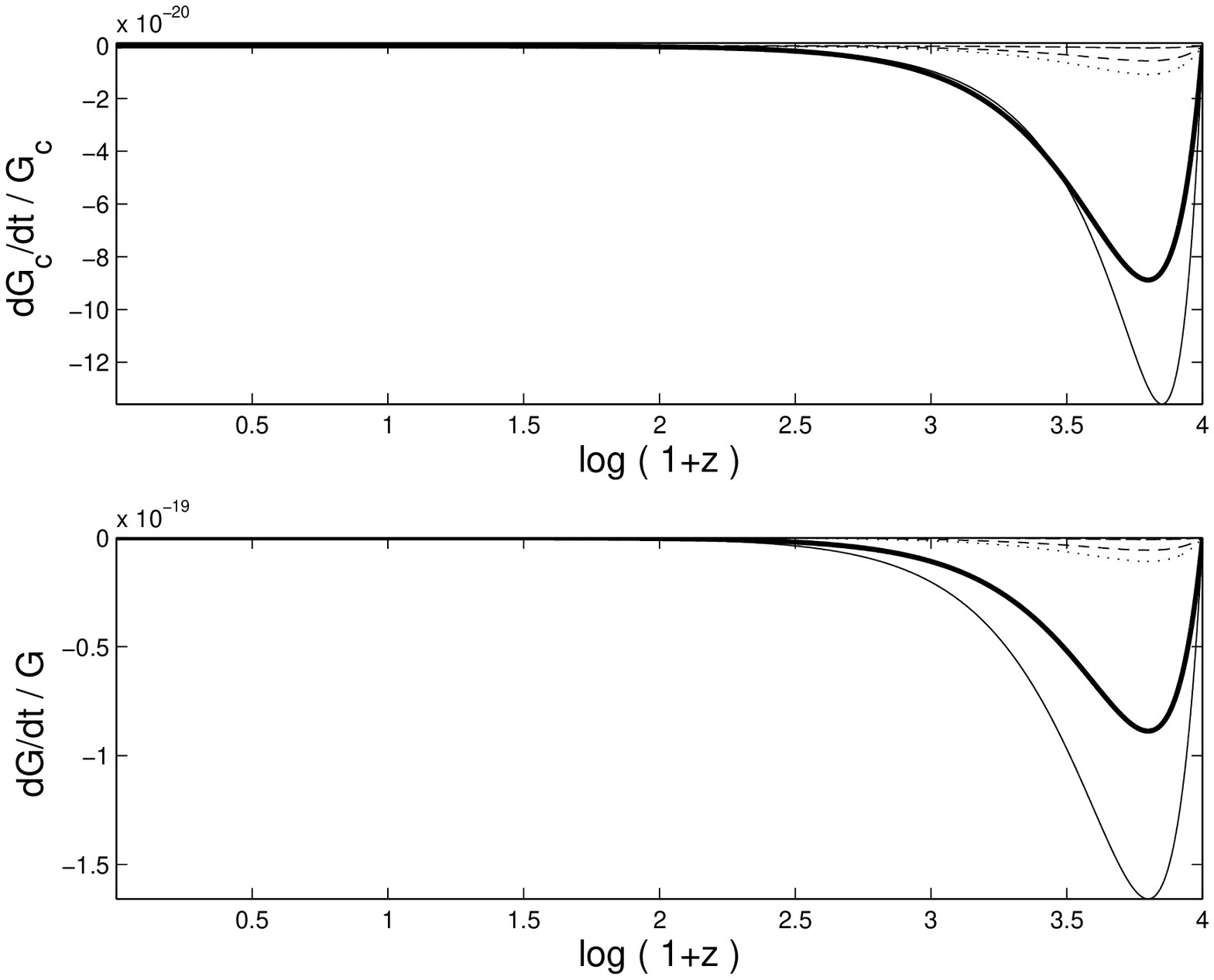,height=6.5cm}
\caption{{\protect {\textit{Evolution of $\dot{G}/G$ as a function of $%
log(1+z)$ for overdensities which virialise at $z=0$ in a $\Lambda $cdm
model. Upper plot: evolution inside the overdensity. Lower plot: evolution
in the background universe. Thick solid line $2\protect\omega +3=4 \times 10^6$,
thin solid line $2\protect\omega +3=B^{2}\protect\phi $,
dashed-line $2\protect\omega +3=2A|1-\frac{\protect\phi }{\protect\phi _{0}}%
|^{-2}$, dash-dotted line $2\protect\omega +3=C|\ln (\frac{\protect\phi }{%
\protect\phi _{0}})|^{-4}$, dotted line $2\protect\omega +3=D|1-(\frac{%
\protect\phi }{\protect\phi _{0}})^{2}|^{-1}$ . Each model is normalised in
order to have $G_{0}=G_{c}(z=0)$ inside the overdensities.}}}}
\label{dotG}
\end{figure}

\begin{figure}%[tbh]
\epsfig{file=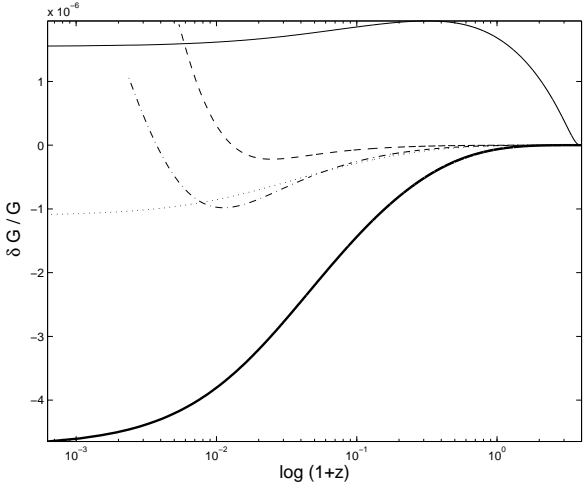,height=6.5cm}
\caption{{\protect {\textit{Evolution of $\protect\delta G/G$ as a
function of $log(1+z)$ for overdensities which virialise at $z=0$ in a $%
\Lambda $cdm model. Thick solid line $2\protect\omega +3=4 \times 10^6$, thin
solid line $2\protect\omega +3=B^{2}\protect\phi ^{2(p+1)}$, dashed-line $2%
\protect\omega +3=2A|1-\frac{\protect\phi }{\protect\phi _{0}}|^{-2}$,
dash-dotted line $2\protect\omega +3=C|\ln (\frac{\protect\phi }{\protect%
\phi _{0}})|^{-4}$, dotted line $2\protect\omega +3=D|1-(\frac{\protect\phi 
}{\protect\phi _{0}})^{2}|^{-1}$ . Each model is normalised in order to have 
$G_{0}=G_{c}(z=0)$ inside the overdensities.}}}}
\label{deltaG}
\end{figure}

The results of our numerical calculations, for a cluster which virialises at 
$z_{v}=0$, are presented in Figures \ref{DeltaG}, \ref{dotG} and \ref{deltaG}%
, respectively. These plots display the evolution of $\dot{G}/G,\Delta G/G,$
and $\delta G/G$ with redshift for Brans-Dicke, the theory of subsection \ref{theory1}, and some other
choices of $\omega (\phi )$ that are specified in the captions.  The
parameters used in generating these plots are $B=0.4$, $C=10^{-16}$,
$D=80$, $A=6 \times 10^{-7}$, $p=2$ and $\omega= 4 \times 10^6$.  These values
were chosen so as to agree with observation and so that structure
formation is not significantly different from that which occurs in
general relativity. 
\begin{figure}%[tbh]
\epsfig{file=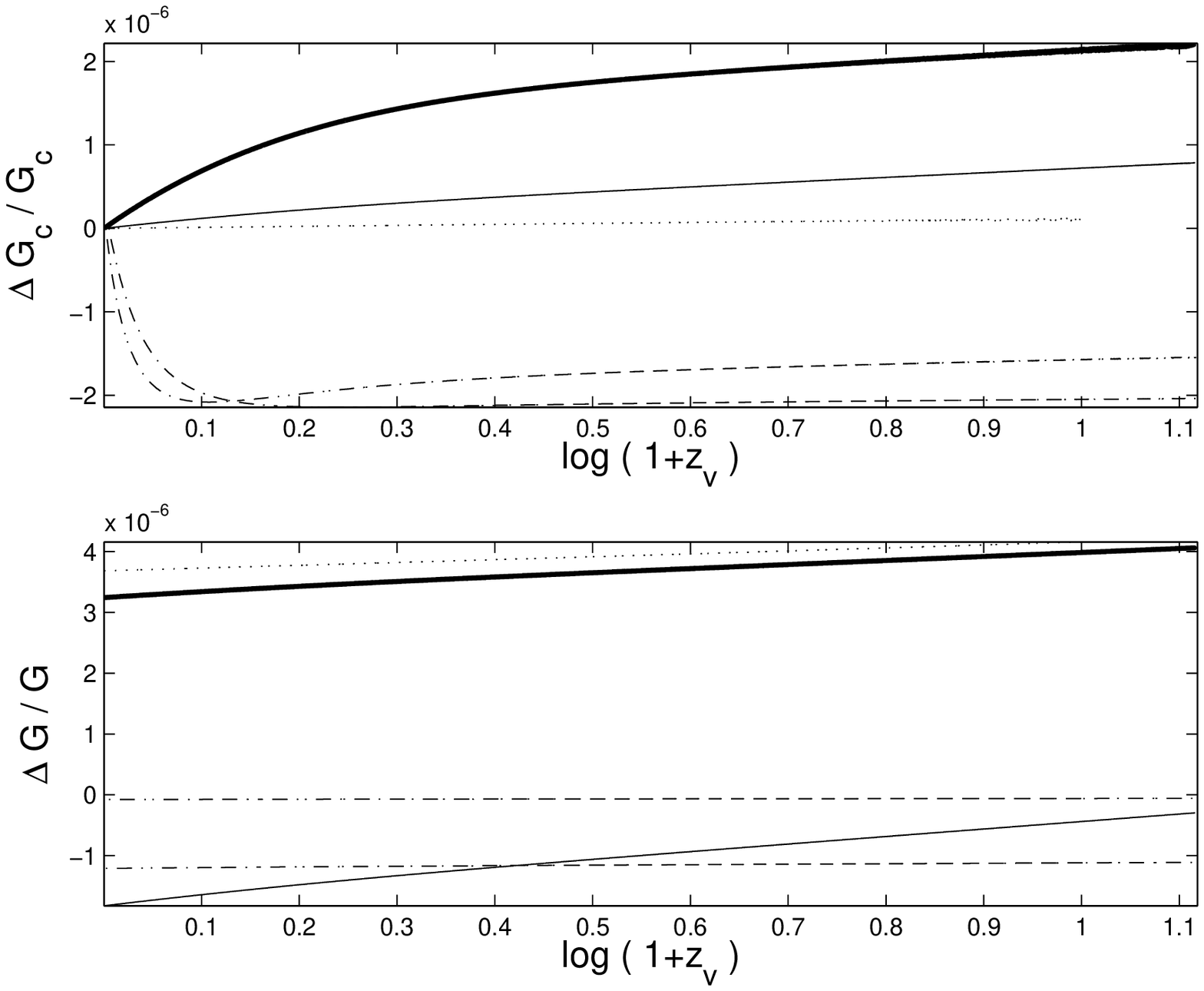,height=6.5cm}
\caption{{\protect {\textit{Values of $\Delta G/G$ as a function of $%
log(1+z_{v})$ at virialisation. Upper plot: evolution inside the
overdensity. Lower plot: evolution in the background universe. Thick solid
line $2\protect\omega +3=4 \times 10^6$; thin solid line $2\protect\omega +3=B^{2}%
\protect\phi $; dashed-line $2\protect\omega +3=2A|1-\frac{\protect%
\phi }{\protect\phi _{0}}|^{-2}$; dash-dotted line $2\protect\omega +3=C|\ln
(\frac{\protect\phi }{\protect\phi _{0}})|^{-4}$; dotted line $2\protect%
\omega +3=D|1-(\frac{\protect\phi }{\protect\phi _{0}})^{2}|^{-1}$ . Each
model is normalised to have $G_{0}=G_{c}(z=0)$ inside the overdensities.}}}}
\label{DeltaGv}
\end{figure}

It is clear from the plots that different scalar-tensor theories lead to
different variations of $G$. The predictions of these models can be quite
diverse. While some models produce higher values of $G$ inside the
overdensity, others produce a lower one. A feature common to all models is
that the value of $G$ and $\dot{G}/G$ inside an overdensity is different
from $G$ and $\dot{G}/G$ in the background universe. The reason for these
differences is that in the non--linear regime, when the overdensity
decouples from the background expansion at turnaround, the field $\phi $
that drives variations in the Newtonian gravitational \textquotedblleft
constant\textquotedblright\ stops feeling the background expansion. After
turnaround, the field inside the overdensity, $\phi _{c}$, deviates from the
field, $\phi $, in the background universe, leading to spatial variations in 
$G$. In reality, such spatial inhomogeneities in the value of $G$ are small: 
$\delta G/G\approx 10^{-6}$, Figure \ref{deltaG}. The time variations of $G$
are even smaller than the spatial inhomogeneities but with a marked
difference between the inside and outside rates of change. We find $\dot{G}%
_{c}/G_{c}\leq 10^{-20}s^{-1}$ inside the clusters and $\dot{G}/G\leq
10^{-19}s^{-1}$ in the background, Figure \ref{dotG}.

\begin{figure}%[tbh]
\epsfig{file=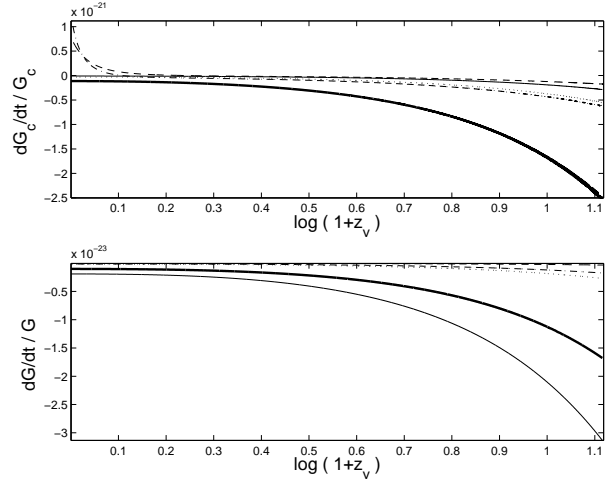,height=6.5cm}
\caption{{\protect {\textit{Values of $\dot{G}/G$ as a function of $%
log(1+z_{v})$ at virialisation. Upper plot: evolution inside the
overdensity. Lower plot: evolution in the background universe. Thick solid
line $2\protect\omega +3=4 \times 10^6$; thin solid line $2\protect\omega +3=B^{2}%
\protect\phi$; dashed-line $2\protect\omega +3=2A|1-\frac{\protect%
\phi }{\protect\phi _{0}}|^{-2}$; dash-dotted line $2\protect\omega +3=C|\ln
(\frac{\protect\phi }{\protect\phi _{0}})|^{-4}$; dotted line $2\protect%
\omega +3=D|1-(\frac{\protect\phi }{\protect\phi _{0}})^{2}|^{-1}$. Each
model is normalised to have $G_{0}=G_{c}(z=0)$ inside the overdensities.}}}}
\label{dotGv}
\end{figure}

\begin{figure}%[tbh]
\epsfig{file=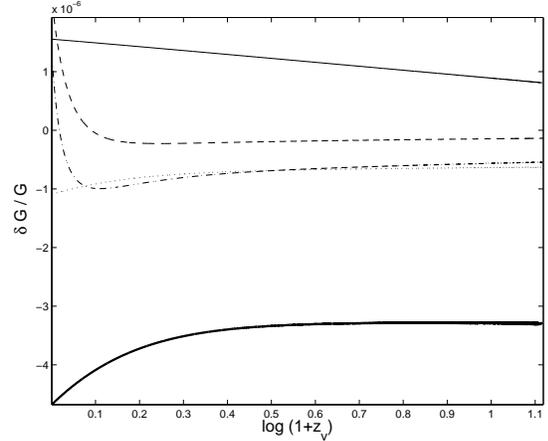,height=6cm}
\caption{{\protect {\textit{Values of $\protect\delta G/G$ as a
function of $log(1+z_{v})$ at virialisation. Thick solid line $2\protect%
\omega +3=4 \times 10^6$; thin solid line $2\protect\omega
+3=B^{2}\protect\phi %
$; dashed-line $2\protect\omega +3=2A|1-\frac{\protect\phi }{%
\protect\phi _{0}}|^{-2}$; dash-dotted line $2\protect\omega +3=C|\ln (\frac{%
\protect\phi }{\protect\phi _{0}})|^{-4}$; dotted line $2\protect\omega %
+3=D|1-(\frac{\protect\phi }{\protect\phi _{0}})^{2}|^{-1}$. Each model is
normalised to have $G_{0}=G_{c}(z=0)$ inside the overdensities.}}}}
\label{deltaGv}
\end{figure}

Figures \ref{DeltaGv}, \ref{dotGv} and \ref{deltaGv} represent the values of 
$\Delta G/G$, $\delta G/G$ and $\dot{G}/G$ at the redshift of virialisation (%
$z_{v}$). Once again the differences between the different scalar-tensor
theories and between $G_{c}(z_{v})$ and $G(z_{v})$ are evident. 

\begin{figure}%[tbh]
\epsfig{file=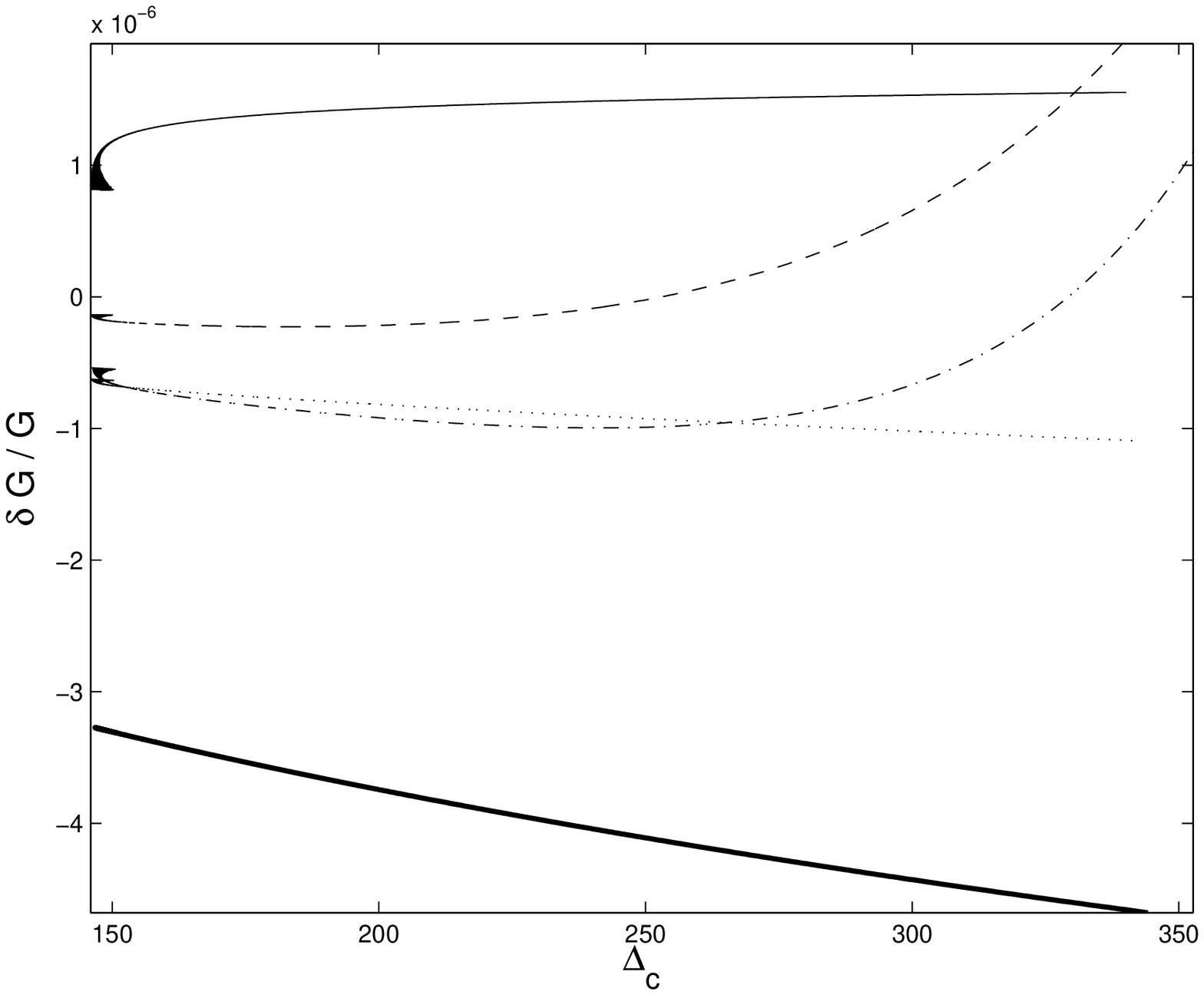,height=6cm}
\caption{{\protect {\textit{Values of $\protect\delta G/G$ as a
function of $\Delta _{c}$ for a $\Lambda $cdm model. Thick-solid line $2%
\protect\omega +3=4 \times 10^6$; solid line $2\protect\omega +3=B^{2}\protect%
\phi $; dashed-line $2\protect\omega +3=2A|1-\frac{\protect\phi }{%
\protect\phi _{0}}|^{-2}$; dash-dotted line $2\protect\omega +3=C|\ln (\frac{%
\protect\phi }{\protect\phi _{0}})|^{-4}$; dotted line $2\protect\omega %
+3=D|1-(\frac{\protect\phi }{\protect\phi _{0}})^{2}|^{-1}$. Each model is
normalised to have $G_{0}=G_{c}(z=0)$ inside the overdensities.}}}}
\label{deltaGrhov}
\end{figure}

Figure \ref{deltaGrhov} shows how the cdm contrast, $\Delta _{c}\equiv \rho
_{cdm}(z_{v})/\rho _{m}(z_{v})$, affects the difference between the value of 
$G$ inside an overdensity and in the cosmological background. (Recall that,
in the Einstein--de Sitter model $\Delta _{\mathrm{c}}\equiv \rho
_{cdm}(z_{v})/\rho _{m}(z_{v})\approx 147$ at virialisation, and $\Delta _{%
\mathrm{c}}\equiv \rho _{cdm}(z_{v})/\rho _{m}(z_{c})\approx 187$ at
collapse). It is interesting to see that different scalar-tensor theories
produce a different dependence. 

From the numerical simulations, we see that a variation of the Newtonian
gravitational constant, \textit{of the order presented here}, does not
affect the predictions of the structure formation models. For instance, the
virialisation radius and the density contrast $\Delta _{c}$ of the
virialised clusters are very similar to the $\Lambda $cdm and standard cdm
models with constant $G$ . Nevertheless, one should point out that different
scalar-tensor theories lead to slightly different results. Besides that,
there is also a dependence on the initial conditions, as with many other
cosmological scalar fields (see e.g. \citep{carsten,mota1,mota2}). Different initial
conditions will lead to slightly different allowed values of $\omega $ and
to different cosmological behaviours of $G$ and $G_{c}$.

\section{Conclusions}

\label{Conclusions}

We have studied the inhomogeneous cosmological evolution of the Newtonian
gravitational 'constant' $G$ within the framework of relativistic
scalar-tensor theories of gravity, of which Brans--Dicke theory is the
simplest and best known case. We began by first exploiting the conformal
equivalence between these theories and general relativity to transform an
existing solution of Einstein's field equations to a new exact spherically
symmetric inhomogeneous vacuum cosmological solution of the Brans--Dicke
field equations. We then present a second spherically symmetric
perfect-fluid solution to the Brans--Dicke field equations which corresponds
to an overdense inhomogeneity in an asymptotically homogenous  and isotropic
expanding cosmological background. These exact solutions have simple enough
form to allow the $G(r,t)$ evolution to be investigated directly and can be
used to model the presence of a Schwarzschild-like mass in an expanding
Brans--Dicke universe that approaches an idealised FRW universe
asymptotically as $r\rightarrow \infty $. It is noted that far from the mass 
$G$ varies very slowly with $r$ and that as $r\rightarrow \infty $ the
variation of $G$ with $r$ is removed altogether and only the cosmological
evolution of $G$ with time is relevant. Close to the mass, the $r$ variation
of $G$ becomes more significant and we see that $G$ $\rightarrow 0$ at small 
$r$. We also note that in the limit $\omega \rightarrow \infty $ all space
and time variation in $G$ is removed from these solutions and general
relativity is recovered.

Next, we increased the complexity of the model by matching together two
vacuum FRW--Brans--Dicke universes of zero and positive curvature on a
spacelike time slice. This results in a simple model for a spherically
symmetric perturbation in the density and curvature that is followed into
the non--linear regime. The different evolutions of $G$ in the two regions
were determined and a comparison was made of the different evolutions of $G$
on spacelike slices of differing time. We see from this toy model that the
value of $G$ does indeed have a different value in regions that have
decoupled from the expanding cosmological background and collapsed, compared
to its value in the background itself. We were then able to repeat this
construction for a more realistic matching of two FRW $\Lambda $cdm
universes of different curvature in an arbitrary scalar--tensor gravity
theory. We followed the development of spherical overdensities through their
expansion, separation from the background, turnaround, and subsequent
collapse. Applying a simple approximation to virialise the collapsing
overdensities, we were then able to study the differences in $G(\vec{x},t)$
between the non-expanding overdensity and the expanding background universe.
We highlight as a special case the simple Brans--Dicke theory but also
present results for other scalar--tensor gravity theories, specified by
their defining coupling function $\omega (\phi )$. Although each theory
predicts a different detailed cosmological evolution of $G$, a feature
common to all of them is the sharp difference between the value and time
evolution of $G$ inside bound overdensities and in the background universe.
These differences will produce spatial inhomogeneities in $G$ with a value
which depends on the scalar--tensor theory used. While some models produce
higher values of $G$ inside the overdensity, others produce a lower one. In
spite of these differences, such spatial inhomogeneities are small $\delta
G/G\approx 10^{-6}$. The differences in the time variations of $G$ were
found, with typically $\dot{G}_{c}/G_{c}\leq 10^{-20}s^{-1}$ inside the
clusters and $\dot{G}/G\leq 10^{-19}s^{-1}$ in the background universe in
the range of theories studied. Such variations in $G$ do not significantly
alter the virialisation radius and the density contrast $\Delta _{c}$ of the
virialised clusters from those in standard $\Lambda $cdm and cdm models with
constant $G$. Nevertheless, different scalar-tensor theories lead to
slightly different results. There is also a dependence on the initial
conditions. Different initial conditions will lead to a different value of $%
\omega $ and to different cosmological behaviours of $G$ and $G_{c}$. Taken
as a whole these analyses show that local observational limits on varying $G$
made within our solar system or Galaxy must be used with caution when
placing constraints upon the allowed cosmological variation of $G$ on
extragalactic scales or in the early universe. The universe is not spatially
homogeneous and, in cosmological models where it can vary, nor is $G$.

\noindent\\

{\large \bf ACKNOWLEDGEMENTS}\\
%
%\section*{Acknowledgments}
%
We would like to thank Andrew Liddle for helpful comments and
suggestions.  DFM is supported by Fundac\~{a}o Ci\^{e}ncia e a Tecnologia. TC is supported
by the PPARC.

\appendix

\section{Inhomogeneous field equations}

\label{appendix}

Substituting (\ref{metric}) into (\ref{field equations}) gives, for a
perfect fluid

\begin{multline}  \label{T11}
\frac{8\pi}{\phi}p = -\left[ 2\frac{\ddot{a}}{a}+\left(\frac{\dot{a}}{a}%
\right)^2 +\frac{\omega}{2} \left( \frac{\dot{\phi}}{\phi} \right)^2 +2\frac{%
\dot{a}}{a}\frac{\dot{\phi}}{\phi} + \frac{\ddot{\phi}}{\phi} \right]
e^{-\nu} \\
+\left[\frac{\mu^{{\prime}^2}}{4} +\frac{\mu^{\prime}\nu^{\prime}}{r} + \frac{%
\mu^{\prime}\nu^{\prime}}{2} -\frac{\omega}{2}\left(\frac{\phi^{\prime}}{\phi%
}\right)^2 \right.\\\left.+\frac{\phi^{\prime}}{\phi} \left(\mu^{\prime}+\frac{\nu^{\prime}%
}{2} +\frac{2}{r}\right)\right] e^{-\mu}a^{-2},
\end{multline}

\begin{multline}  \label{T22}
\frac{8\pi}{\phi}p=-\left[2\frac{\ddot{a}}{a}+\left(\frac{\dot{a}}{a}
\right)^2 +\frac{\omega}{2} \left(\frac{\dot{\phi}}{\phi} \right)^2+2\frac{%
\dot{a}}{a} \frac{\dot{\phi}}{\phi} + \frac{\ddot{\phi}}{\phi} \right]
e^{-\nu} \\
+\left[ \frac{\mu^{\prime\prime}\nu^{\prime\prime}}{2} + \frac{\nu^{{\prime}^2}%
}{4} + \frac{\mu^{\prime}+\nu^{\prime}}{2r} +\frac{\omega}{2} \left(\frac{%
\phi^{\prime}}{\phi} \right)^2 + \right.\\\left.\frac{\phi^{\prime\prime}}{\phi} + \left(%
\frac{\nu^{\prime}}{2} +\frac{1}{r}\right) \frac{\phi^{\prime}}{\phi} \right]
e^{-\mu}a^{-2},
\end{multline}

and

\begin{multline}  \label{T44}
-\frac{8\pi}{\phi} \rho = -\left[3\left(\frac{\dot{a}}{a}\right)^2 - \frac{%
\omega}{2} \left(\frac{\dot{\phi}}{\phi}\right)^2+3\frac{\dot{a}}{a}\frac{%
\dot{\phi}}{\phi}\right] e^{-\nu} \\
+\left[ \mu^{\prime\prime}+\frac{\mu^{{\prime}^2}}{4} + \frac{2\mu^{\prime}}{r}
+\frac{\omega}{2} \left(\frac{\phi^{\prime}}{\phi} \right)^2 \right.\\\left.+\frac{%
\phi^{\prime\prime}}{\phi} +\left(\frac{\nu^{\prime}}{2} +\frac{1}{r}\right)%
\frac{\phi^{\prime}}{\phi}\right] e^{-\mu} a^{-2}
\end{multline}

as the $T^{11}$, $T^{22}$ and $T^{00}$ equations, respectively. The
propagation equation (\ref{box phi}) now becomes

\begin{multline}  \label{wave}
\frac{8\pi(\rho-3p)}{(2\omega+3)\phi} = \left[\frac{\ddot{\phi}}{\phi} + 3%
\frac{\dot{a}}{a}\frac{\dot{\phi}}{\phi}\right] e^{-\nu} \\-\left[\frac{%
\phi^{\prime\prime}}{\phi} + \frac{(\mu^{\prime}+\nu^{\prime})}{2} \frac{%
\phi^{\prime}}{\phi} + \frac{2}{r} \frac{\phi^{\prime}}{\phi}\right]%
e^{-\mu}a^{-2}.
\end{multline}

and the only other non-trivial field equation is that for $T^{10}$;

\begin{equation}  \label{T14}
\nu^{\prime}\frac{\dot{a}}{a} - \omega\frac{\dot{\phi}}{\phi}\frac{%
\phi^{\prime}}{\phi} - \frac{\dot{\phi}^{\prime}}{\phi} +\frac{\dot{a}}{a}%
\frac{\phi^{\prime}}{\phi} + \frac{\nu^{\prime}}{2} \frac{\dot{\phi}}{\phi}%
=0.
\end{equation}

We now assume $\phi$ is of the form $\phi(r,t)=\phi(r)\phi(t)$ and look for
solutions to the set of equations

\begin{multline}  \label{S11}
\frac{\mu^{{\prime}^2}}{4} +\frac{\mu^{\prime}\nu^{\prime}}{r} + \frac{%
\mu^{\prime}\nu^{\prime}}{2} \\=\frac{\omega(r)}{2}\left(\frac{\phi^{\prime}(r)%
}{\phi(r)}\right)^2 -\frac{\phi^{\prime}(r)}{\phi(r)} \left(\mu^{\prime}+%
\frac{\nu^{\prime}}{2} +\frac{2}{r}\right),
\end{multline}

\begin{multline}  \label{S22}
\frac{\mu^{\prime\prime}\nu^{\prime\prime}}{2} + \frac{\nu^{{\prime}^2}}{4} + 
\frac{\mu^{\prime}+\nu^{\prime}}{2r} \\=-\frac{\omega}{2} \left(\frac{%
\phi^{\prime}(r)}{\phi(r)} \right)^2 - \frac{\phi^{\prime\prime}(r)}{\phi(r)}
- \left(\frac{\nu^{\prime}}{2} +\frac{1}{r}\right) \frac{\phi^{\prime}(r)}{%
\phi(r)},
\end{multline}

\begin{multline}  \label{S44}
\mu^{\prime\prime}+\frac{\mu^{{\prime}^2}}{4} + \frac{2\mu^{\prime}}{r} \\=-%
\frac{\omega}{2} \left(\frac{\phi^{\prime}(r)}{\phi(r)} \right)^2 -\frac{%
\phi^{\prime\prime}(r)}{\phi(r)} -\left(\frac{\nu^{\prime}}{2} +\frac{1}{r}%
\right)\frac{\phi^{\prime}(r)}{\phi(r)},
\end{multline}

and

\begin{equation}  \label{Swave}
\frac{\phi^{\prime\prime}(r)}{\phi(r)} + \frac{(\mu^{\prime}+\nu^{\prime})}{2%
} \frac{\phi^{\prime}(r)}{\phi(r)} + \frac{2}{r} \frac{\phi^{\prime}(r)}{%
\phi(r)}=0.
\end{equation}

Such solutions are given by \citep{Nariai68} as

\begin{equation}
e^{\nu }=\left( \frac{1-\frac{c}{2kr}}{1+\frac{c}{2kr}}\right) ^{2k},
\label{ev2}
\end{equation}

\begin{equation}
e^{\mu }=\left( 1+\frac{c}{2kr}\right) ^{4}\left( \frac{1-\frac{c}{2kr}}{1+%
\frac{c}{2kr}}\right) ^{2(k-1)(k+2)/k},  \label{eu2}
\end{equation}

and

\begin{equation}
\phi (r)=\phi _{0}\left( \frac{1-\frac{c}{2kr}}{1+\frac{c}{2kr}}\right)
^{-2(k^{2}-1)/k}  \label{phir2}
\end{equation}

where $k=\sqrt{\frac{4+2\omega}{3+2\omega}}$. For $e^{\nu}$ and $e^{\mu}$ of
this form, equations (\ref{T11}), (\ref{T44}) and (\ref{wave}) become

\begin{equation}  \label{t11}
\frac{8\pi(\rho e^{\nu}-3p e^{\nu})}{(2\omega+3)\phi(t)} = \frac{\ddot{%
\phi(t)}}{\phi(t)} + 3\frac{\dot{a}}{a}\frac{\dot{\phi(t)}}{\phi(t)},
\end{equation}

\begin{equation}  \label{t44}
\frac{8\pi}{\phi(t)} \rho e^{\nu} = 3\left(\frac{\dot{a}}{a}\right)^2 - 
\frac{\omega}{2} \left(\frac{\dot{\phi(t)}}{\phi(t)}\right)^2+3\frac{\dot{a}%
}{a}\frac{\dot{\phi(t)}}{\phi(t)},
\end{equation}

and

\begin{multline}  \label{wave2}
-\frac{8\pi}{3\phi(t)}\frac{(3\omega pe^{\nu} + 3\rho e^{\nu} + \omega\rho
e^{\nu})}{(2\omega+3)}\\=\frac{\ddot{a}}{a} -\frac{\dot{a}}{a}\frac{\dot{%
\phi(t)}}{\phi(t)} + \frac{\omega}{3}\left(\frac{\dot{\phi(t)}}{\phi(t)}%
\right)^2.
\end{multline}

Where (\ref{wave2}) was obtained by substituting (\ref{T11}) and (\ref{T44})
into (\ref{wave}) and discarding the terms involving $r$ derivatives, as
these now sum to zero. We see that (\ref{t11}), (\ref{t44}) and (\ref{wave2}%
) are simply (\ref{Nariai1}), (\ref{Nariai2}) and (\ref{Friedmann}) with $%
k=0 $, $p_{FRW}=pe^{\nu }$ and $\rho _{FRW}=\rho e^{\nu }$, where subscript $%
_{FRW}$ denotes a quantity derived from the field equations using the FRW
metric. We also have, from $T_{;\nu }^{\mu \nu }=0$, that

\begin{equation}  \label{fluid2}
\frac{d}{dt}(\rho e^{\nu}) + 3H(\rho e^{\nu} +p e^{\nu}) = 0.
\end{equation}

Looking for solutions of the form $a\propto t^{x}$ and $\phi (t)\propto t^{y}
$ gives, on substitution into (\ref{t11}), the relation $y=2-3x\gamma $
(assuming an equation of state for the Universe of the form $p=(\gamma
-1)\rho $). Using (\ref{t11}), (\ref{t44}), and this relation then gives the
solutions

\begin{equation}  \label{a2}
a(t)=a_0\left(\frac{t}{t_0}\right)^{\frac{2\omega(2-\gamma)+2}{%
3\omega\gamma(2-\gamma)+4}},
\end{equation}

and

\begin{equation}  \label{phi(t)2}
\phi(t)=\phi_0\left(\frac{t}{t_0}\right)^{\frac{2(4-3\gamma)}{%
3\omega\gamma(2-\gamma)+4}}.
\end{equation}

These are exactly the same as would be expected for the scale factor and the
BD field in a flat FRW Universe \citep{nar}. The form of $\rho$ is then given
by (\ref{fluid2}) and (\ref{ev2}) as

\begin{equation}
\rho (r,t)=\rho _{0}\left( \frac{a_{0}}{a(t)}\right) ^{3\gamma }\left( \frac{%
1-\frac{c}{2kr}}{1+\frac{c}{2kr}}\right) ^{-2k},  \label{rho(rt)2}
\end{equation}

and $\phi(r,t)$ is given as

\begin{equation}
\phi (r,t)=\phi _{0}\left( \frac{t}{t_{0}}\right) ^{\frac{2(4-3\gamma )}{%
3\omega \gamma (2-\gamma )+4}}\left( \frac{1-\frac{c}{2kr}}{1+\frac{c}{2kr}}%
\right) ^{-2(k^{2}-1)/k}.  \label{phi(rt)2}
\end{equation}

\label{lastpage}
\end{document}